\documentclass[final,3p,times,authoryear]{elsarticle}

\usepackage{amssymb}
\usepackage{amsmath}
\usepackage{float}
\usepackage{graphicx}
\usepackage{xcolor}
\usepackage{draftwatermark} 
\usepackage[scale=15,opacity=0.2,color=gray]{background}

\journal{Journal of the Mechanics and Physics of Solids}

\begin{document}

\begin{frontmatter}



\title{Preprint\fnref{note3}: Towards Interfacing Dark-Field X-ray Microscopy to Dislocation Dynamics Modeling} 

\author[1]{Axel Henningsson \corref{cor1}\fnref{note1}}
\author[2]{Sina Borgi\fnref{note1}}
\author[1]{Grethe Winther}
\author[3]{Anter El-Azab}
\author[2]{Henning Friis Poulsen}

\cortext[cor1]{Corresponding author: naxhe@dtu.dk}
\fntext[note1]{These authors contributed equally to the presented work}
\fntext[note3]{This is a preprint. Please cite the published journal version whenever possible. }
\affiliation[1]{organization={Department of Mechanical Engineering, Technical University of Denmark}, 
          addressline={}, 
          city={Kgs. Lyngby}, 
          postcode={2800}, 
          state={}, 
          country={Denmark}}
          
\affiliation[2]{organization={Department of Physics, Technical University of Denmark}, 
          addressline={}, 
          city={Kgs. Lyngby}, 
          postcode={2800}, 
          state={}, 
          country={Denmark}}

\affiliation[3]{organization={School of Nuclear Engineering and School of Materials Engineering, Purdue University}, 
          addressline={}, 
          city={West Lafayette}, 
          postcode={47907}, 
          state={IN}, 
          country={USA}}

\begin{abstract}
Deformation gradient tensor fields are reconstructed in three dimensions (mapping all 9 tensor components) using synthetic Dark-Field X-ray Microscopy data. Owing to the unique properties of the microscope, our results imply that the evolution of deformation fields can now be imaged non-destructively, in situ, and within deeply embedded crystalline elements. The derived regression framework and sampling scheme operate under the kinematic diffraction approximation and are well-suited for studying microstructure evolution during plastic deformation. We derive the deformation conditions under which diffraction vectors extracted from DFXM images can be uniquely associated to the deformation gradient tensor field of the sample. The analysis concludes that the deformation gradient tensor field must vary linearly over line segments defined by the X-ray beam width and the diffracted ray path. The proposed algorithms are validated against numerical simulations for realistic noise levels. Reconstructions of a simulated single straight-edge dislocation show that the Burgers vector components can be recovered with an error of \(<\)2\%. The mean absolute error of the reconstructed elastic distortion field was found to be \(<\)\(10^{-6}\). By taking the curl of the elastic distortion field, local dislocation densities are derived, yielding a reconstructed dislocation core position with sub-pixel accuracy. The significance of directly measuring the elastic distortion and the dislocation density tensor fields is discussed in the context of continuum theory of dislocations. Such measurements can also be interfaced with continuum dislocation dynamics by providing data that can guide the development and validation, thus extending the relevant models to finite strain regimes.
\end{abstract}


\end{frontmatter}

\backgroundsetup{contents={\sffamily Preprint}}
\section{Introduction}\label{sec:Introduction}
Crystallographic line defects known as dislocations are the carriers of plastic deformation in metals. The significance of dislocations in this context was first realized in the 1930s, when Polanyi, Taylor, and Orowan recognized that the plastic flow of metals at (then) unpredicted, and very low, shear stresses could be explained by these defects \citep{Hirth1985}. Since then, the dynamics of dislocations have been firmly established as central to the phenomenon of metal plasticity. It is now widely recognized that the macroscopic properties of crystalline materials are determined to a large extent by the behavior of their constituent defects \citep{Hussein2012}. The coupling between dislocations and plasticity manifests through dislocation interactions and multiplication, leading to work-hardening and the self-organization of dislocations into patterns. However, the underlying mechanisms and driving forces remain elusive, largely due to the lack of experimental techniques capable of visualizing the process and resolving the associated mechanical fields.

Over the past three decades, Discrete Dislocation Dynamics (DDD) models have evolved in tandem with the advancement of computers \citep{Devincre1997, Cai2004, Mohamed2015}. These models focus on simulating individual dislocations and their interactions, with the feasible model domain volume and dislocation density directly constrained by the available computational performance. As deformation increases, dislocation densities can reach prohibitively high levels, with values as large as 10\(^{12}\) cm\(^{-2}\) observed in heavily deformed metals \citep{Ohring1995}. Consequently, DDD serves as a powerful tool for studying strain in regimes of \(<\)1\%. At larger deformations, a complementary framework, Continuum Dislocation Dynamics (CDD), becomes necessary. Unlike DDD, CDD operates from a continuum perspective, enabling the dislocation density tensor field, \(\boldsymbol{\alpha}\), and its constituents, to vary continuously over the model domain \citep{El-Azab2000, A2Cermelli2001, Acharya2001, Hochrainer2009, A3Starkey2020, A5Vivekanandan2021, A4Starkey2022}.

Regardless of the magnitude of dislocation density, state-of-the-art DDD and CDD models can provide dynamic representations of crystalline structures evolving within deeply embedded three-dimensional spaces. As both approaches rely on the description of crystalline structures through collections of higher-order tensor fields, comprehensive experimental validation remains a significant challenge to this day.

Dark-Field X-ray Microscopy (DFXM) is a new full-field imaging technique that enables non-destructive 3D mapping of deeply embedded crystalline elements, such as grains or domains \citep{Simons2015, Simons2018, Bucsek2019, Mavrikakis2019,
Poulsen2020}, and dislocations \citep{Jakobsen2019, Simons2019, Dresselhaus-Marais2021, Yildirim2023}, in bulk samples. With a spatial resolution of \(\sim\)100 nm and a working distance in the centimeter range, DFXM offers unique capabilities for in situ investigation of metal plasticity. Unlike traditional techniques such as Transmission Electron Microscopy (TEM), which provide atomic-scale resolution but are limited to thin foils, DFXM can image embedded volumes of up to 1000 \(\mu\)m\(^3\) and capture dynamic processes in bulk materials. By additionally offering strain resolution of \(10^{-4}\) or better DFXM is well situated for in situ studies of bulk representative deformation.

The first dedicated dark-field X-ray microscope was recently revamped and moved to the ID03 beamline of the European Synchrotron Radiation Facility where it was originally installed at the ID06 beamline \citep{Kutsal2019}. Several tests have also been performed
at X-ray Free Electron Laser (XFEL) facilities \citep{Dresselhaus-Marais2023, Holstad2023}. Considering the time resolution of around 200 ms at synchrotrons and 1 ps at XFELs, DFXM offers a unique opportunity to validate the predicted dynamics of DDD and CDD models. Indeed, for low dislocation densities, mapping and tracking of individual dislocations as a function of annealing was recently demonstrated by \cite{Dresselhaus-Marais2021}. To fully interface DFXM with DDD simulations, however, the accompanying spatial tensor fields, describing the lattice deformation must be measured. Such tensor fields are still lacking. This issue becomes even more critical for CDD models, where high dislocation densities make the mapping of individual dislocation lines, and their associated Burgers vectors and slip systems, unfeasible. 

Seeking to bridge this gap, we present a framework for extracting quantitative deformation tensor fields from DFXM data. Central to our regression scheme is a type of feature extraction from the DFXM image data which massively reduces the dimensionality of the raw intensity image stack to a single mean diffraction response per detector pixel - represented by three angular coordinates of the microscope, \(\phi, \chi\) and \( \Delta\theta\). In DFXM, this data reduction step can be performed by analyzing the variation in diffracted intensity at a given detector pixel as a function of systematically perturbing \(\phi, \chi\) and \( \Delta\theta\). Importantly, our work is limited to elastic scattering in which a geometrical optics approximation of diffraction is acceptable, see \cite{Poulsen2017, Poulsen2021, Borgi2024}. Assuming that a compound refractive lens is being used as an objective we show how local diffraction vectors (existing in reciprocal space) can be reconstructed from the extracted mean diffraction angles in \(\phi, \chi\) and \( \Delta\theta\) (existing in angular space). This allows us to provide a link between diffraction conditions and mechanical deformation theory. We then demonstrate, through numerical simulations, that the full 3D elastic deformation gradient tensor field, \(\boldsymbol{F}\), can be reconstructed from stacks of DFXM intensity images in this setting. Importantly, the method we present in this work represent a general framework to extract lattice deformation fields in an embedded crystalline element from DFXM data. In fact, the developed method is fully agnostic to the source of variation in the elastic crystal deformation gradient tensor field, \(\boldsymbol{F}\), which defines the local mapping between undeformed and deformed lattice states. This broad applicability also extends to the derivation of tensor metrics crucial specifically to DDD and CDD, including stress, strain, elastic distortion, local orientation, and, through differentiation, dislocation density. To showcase this emerging interface between DFXM and dislocation dynamics models, we demonstrate our method on a numerical phantom featuring a straight edge dislocation.

Beyond interfacing DFXM data with dislocation dynamics models, we envision that extracting the dislocation density tensor field from experimental DFXM data will in itself be a crucial tool for addressing outstanding scientific questions about the origins of patterning and work-hardening. These include the composition of dislocation boundaries, the associated active slip systems, and the nature of the long-range stress fields around dislocation boundaries. For the latter, several competing hypotheses exist, ranging from the Low Energy Dislocation Structures (LEDS) principle \citep{Wilsdorf1999}, which suggests that dislocations shield each other's stress fields, leading to patterns free of long-range stresses, to models proposing significant internal stresses, such as the composite model, which describes stress partitioning between dislocation boundaries and dislocation-free domains \citep{Mughrabi2002}.

\begin{figure}[H]
    \centering
    \includegraphics[width=0.7\linewidth]{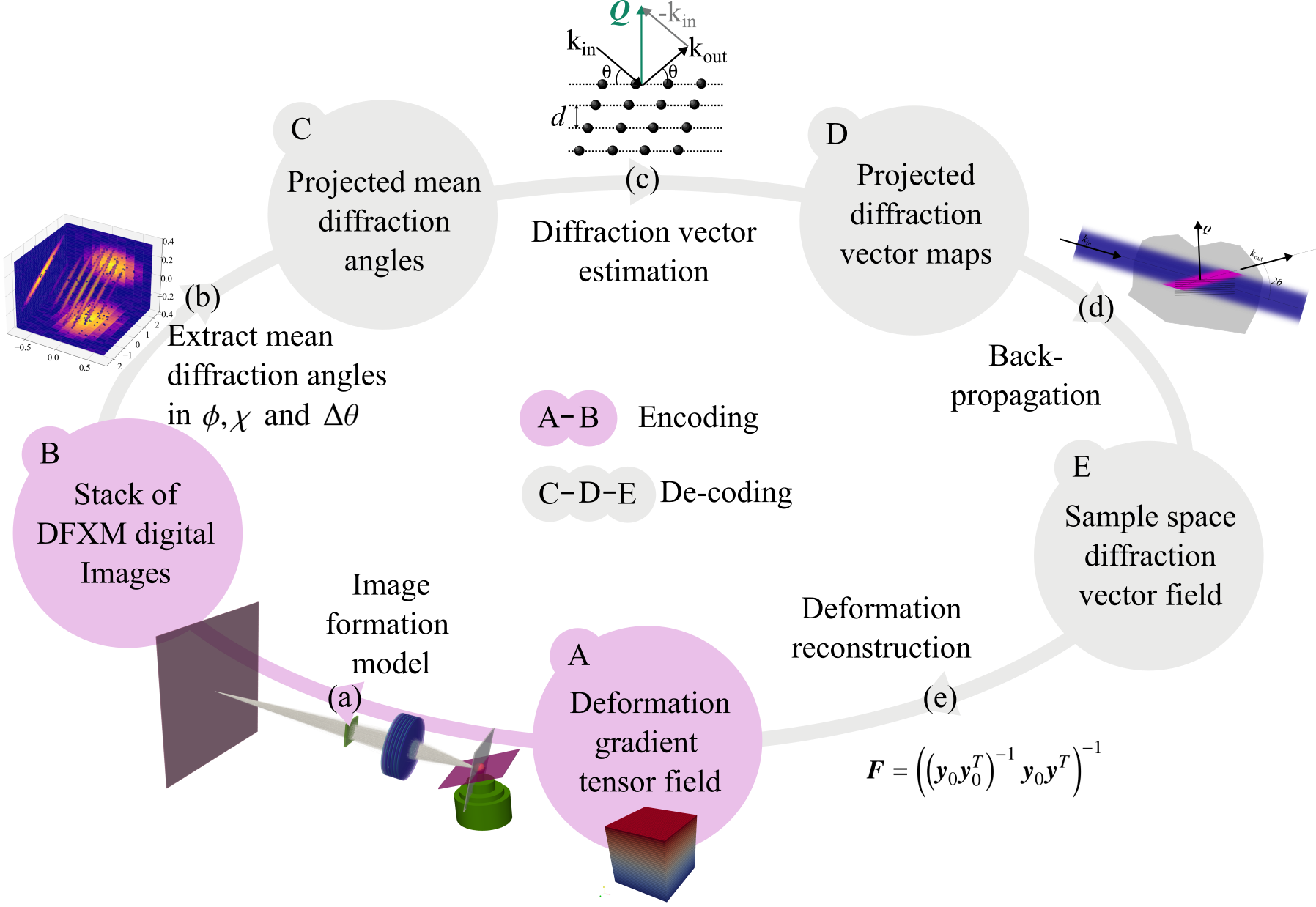}
    \caption{Overview of presented framework. A deformation gradient tensor field (A) is input to a numerical forward model of diffraction (a) to produce a stack of DFXM images (B) encoding the deformation field. The 2D digital images (from multiple reflections) are then compiled to give the variation in diffracted intensity at a given detector pixel as a function of systematically perturbing the microscope geometry (b). The local diffraction response in (b) is then analyzed for its mean value resulting in a mean diffraction angle in \(\phi, \chi\) and \( \Delta\theta\) per detector pixel and reflection (C). By analyzing the spatial map of mean diffraction angle in \(\phi, \chi\) and \( \Delta\theta\) (c) corresponding projected maps of diffraction vectors, \(\boldsymbol{Q}\), can be established (D), and by back-propagating (d) these maps from the detector image space to the sample real space each voxel in the sample is associated to a set of \(\boldsymbol{Q}\)-vectors (E). Finally, the deformation field, \(\boldsymbol{F}\), is reconstructed (e) from the diffraction vector field providing again a deformation gradient tensor field.}
    \label{fig:scheme}
\end{figure}

The framework presented in this paper is validated through a new numerical model, used to generate synthetic DFXM images from an input deformation gradient tensor field (Figure \ref{fig:scheme}A-B). The associated regression scheme, used to analyze the diffraction signal and reconstruct the deformation gradient tensor field is discussed in detail (Figure \ref{fig:scheme}C-E). By de-coding the DFXM diffraction signal in the proposed way we allow for the integration of experimental DFXM movies with CDD and DDD simulations, enabling full-field, full-tensorial characterization of plastic deformation in bulk metals, providing, in this way, a unified tool for studying dislocation dynamics and plasticity at unprecedented resolution.
The paper is structured as follows: in section \ref{sec:Diffraction from a deformed crystal} we connect the theory of kinematical scattering to the lattice deformation field. Next, we introduce the DFXM microscope in section \ref{sec:The Dark-Field X-ray Microscope} and give the necessary measurement conditions of our framework. The five steps of analysis in Figure \ref{fig:scheme}(a)-(e) are then presented in sections \ref{sec:Image formation} and \ref{sec:Regression scheme for determining the deformation gradient tensor field} providing details on the image formation model used to generate synthetic DFXM images and presenting our regression scheme. Section \ref{sec:Numerical Example} provides a validation of the presented theory using numerical simulations. Finally, we discuss our findings and close the paper in section \ref{sec:Discussion}.

\section{Diffraction from a deformed crystal}\label{sec:Diffraction from a deformed crystal}
A connection between the diffraction conditions for a crystalline element and the corresponding lattice deformation field was first formulated by \cite{Bernier2011} in the context of (lensless) far-field high-energy diffraction microscopy. The same connection was later re-established in \cite{Poulsen2021}, then in the context of DFXM. Building on these previous works, we introduce the unit cell matrix \(\boldsymbol{C}\in\mathbb{R}^{3\times 3}\) as
\begin{equation}
    \boldsymbol{C} = \begin{bmatrix}
        \boldsymbol{a}_1 & \boldsymbol{a}_2 & \boldsymbol{a}_3
    \end{bmatrix},
    \label{eq:unit_cell_matrix}
\end{equation}
where \(\boldsymbol{a}_i\in\mathbb{R}^3\) are vectors that define the lattice unit cell parallelepiped in real space. Here (and throughout this paper) when no reference to a specific coordinate systems is given, it can be assumed that an arbitrary but consistent frame of reference is to be deployed. Next, we define the elastic diffraction vector as the wavevector difference
\begin{equation}
    \boldsymbol{Q} = \boldsymbol{k} - \boldsymbol{k'},
    \label{eq:elastic_diffraction_vector}
\end{equation}
where \(\boldsymbol{k}\) and \(\boldsymbol{k'}\) are the wavevectors of the incoming and outgoing X-ray beams, respectively. With these definitions, the Laue equations are
\begin{equation}
    \boldsymbol{C}^T\boldsymbol{Q} = 2 \pi \boldsymbol{Q}_{hkl},
    \label{eq:laue}
\end{equation}
where \(\boldsymbol{Q}_{hkl}\in\mathbb{R}^3\) contains (integer) Miller indices \((h,k,l)\). In a DFXM setting, \(\boldsymbol{Q}_{hkl}\) can be treated as a known constant (derived from the average crystal orientation and lattice symmetry).

Considering a compatible displacement field, \(\boldsymbol{u}\in\mathbb{R}^{3}\), we can define a corresponding deformation gradient tensor field, \(\boldsymbol{F}\in\mathbb{R}^{3\times 3}\), as 
\begin{equation}
    \boldsymbol{F} = \nabla \boldsymbol{u} + \boldsymbol{I},
    \label{eq:deformation_gradient}
\end{equation}
where \(\nabla \boldsymbol{u}\in\mathbb{R}^{3\times 3}\) is the referential gradient of the displacement field and \(\boldsymbol{I}\in\mathbb{R}^{3\times 3}\) is the identity tensor. Let \(\boldsymbol{x} ,  \boldsymbol{x}^{(0)} \in\mathbb{R}^3\) denote the coordinates of the material points in the deformed state and the initial state of the material, respectively. The deformation gradient tensor in equation \eqref{eq:deformation_gradient} satisfies the differential form
\begin{equation}
    d \boldsymbol{x} = \boldsymbol{F} d \boldsymbol{x}^{(0)},
    \label{eq:deformation}
\end{equation}
where \(d \boldsymbol{x}^{(0)}\in\mathbb{R}^3\) denotes an incremental vector line segment in the reference configuration and \(d \boldsymbol{x}\in\mathbb{R}^3\) is a corresponding deformed line segment. For a general deformed state, the deformation gradient tensor is expressed as a multiplicative decomposition of the elastic, $\boldsymbol{F}^{e}$ and the plastic $\boldsymbol{F}^{p}$, deformation gradient tensors. This decomposition takes the form: $\boldsymbol{F}=\boldsymbol{F}^{e}\boldsymbol{F}^{p}$. The elastic and plastic deformation gradient tensors themselves can be written in the form of $\boldsymbol{F}^{e}=\boldsymbol{\beta}^{e}+\boldsymbol{I}$ and $\boldsymbol{F}^{p}=\boldsymbol{\beta}^{p}+\boldsymbol{I}$ where \(\boldsymbol{\beta}^e,\boldsymbol{\beta}^p\in\mathbb{R}^{3\times 3}\) are the elastic and plastic distortion tensors, respectively. In the case of infinitesimal deformation, the deformation gradient tensor can be expanded as follows,
\begin{equation}
    \boldsymbol{F} = \boldsymbol{\beta}^{e} +  \boldsymbol{\beta}^{p}+\boldsymbol{I} = \nabla \boldsymbol{u} + \boldsymbol{I}. 
\end{equation}
The above formalism of crystal distortion is suitable for crystals containing defects, the introduction of which is expressed by the plastic part of the distortion; see Refs \citep{El-Azab2000, A2Cermelli2001, Acharya2001, Hochrainer2009, A3Starkey2020, A5Vivekanandan2021, A4Starkey2022} for more details. Here, elastic distortions, \(\boldsymbol{F}^{e}\) and \(\boldsymbol{\beta}^{e}\), account for both mechanical boundary-induced deformation and lattice distortion caused by underlying defects. Those defects arise due to the non-uniformity of the plastic distortion \(\boldsymbol{F}^{p}\) or \(\boldsymbol{\beta}^{p}\). As pointed out by \cite{Henningsson2023}, in the context of diffraction, the plastic part of the deformation is not observable. To understand this we note that while the Laue equations \eqref{eq:laue} are sensitive to perturbations in the cell matrix, \(\boldsymbol{C}\), the cell matrix itself does not denote a set of fixed lattice motifs. Instead the cell matrix is a representation of the local average unit cell of the crystal lattice. To exemplify; in the presence of pure slip, when lattice sites are systematically displaced, the local cell matrix nevertheless remains unchanged. Considering equations \eqref{eq:deformation} and \eqref{eq:unit_cell_matrix} in this context we therefore find that
\begin{equation}
    \boldsymbol{C} = \boldsymbol{F}^{e}\boldsymbol{C}^{(0)}.
    \label{eq:deformation_lattice}
\end{equation}
For ease of notation, we shall denote \(\boldsymbol{F}^{e}\) by \(\boldsymbol{F}\) in the sequel, and refer to this quantity simply as \textit{the deformation gradient tensor}. Such a notation also complies with what is commonly deployed in the context of diffraction measurements, c.f \cite{Bernier2011, Poulsen2021, Henningsson2023, Borgi2024}. Note that even though the result of \eqref{eq:deformation_lattice} is conditioned only on \(\boldsymbol{F}^{e}\), our theory is still applicable to systems where the plastic distortions \(\boldsymbol{F}^{p}\) and \(\boldsymbol{\beta}^{p}\) are non-zero.

By comparing equations \eqref{eq:laue} and \eqref{eq:deformation_lattice} it is clear that the deformation gradient tensor \(\boldsymbol{F}\) is directly related to the diffraction vector \(\boldsymbol{Q}\). By insertion we find that
\begin{equation}
\big(\boldsymbol{C}^{(0)}\big)^T\boldsymbol{F}^T\boldsymbol{Q} = 2 \pi \boldsymbol{Q}_{hkl}.
    \label{eq:laue_F}
\end{equation}
Reconstructing \(\boldsymbol{F}\) from equation \eqref{eq:laue_F} using DFXM data is the central goal of this research. From the deformation gradient tensor field, \(\boldsymbol{F}\), the elastic distortion tensor \(\boldsymbol{\beta}^e\in\mathbb{R}^{3\times 3}\) (c.f \cite{ElAzab2020}) can be defined as
\begin{equation}
    \boldsymbol{\beta}^e = \boldsymbol{F} - \boldsymbol{I},
    \label{eq:elastic_distortion}
\end{equation}
such that equation \eqref{eq:laue_F} establishes a direct connection between the Laue equations and CDD. To again simplify the notation we shall denote \(\boldsymbol{\beta}^e\) by \(\boldsymbol{\beta}\) in the sequel and refer to this quantity as \textit{the elastic distortion tensor}. 

Another central quantity in the CDD frameworks is the dislocation density tensor field \(\boldsymbol{\alpha}\in\mathbb{R}^{3\times 3}\) \citep{Nye1953, A1Kroner1981} which is defined as the curl of the elastic distortion tensor field
\begin{equation}
    \boldsymbol{\alpha} = \nabla \times \boldsymbol{\beta}.
    \label{eq:dislocation_density}
\end{equation}
The dislocation density tensor defined by equation \eqref{eq:dislocation_density} plays a central role  in the current work. It is the main representation of the dislocations retained in the crystal following plastic deformation. As we shall see later, when \(\boldsymbol{\alpha}\) can be resolved down to the spacing between dislocations, the dislocation density tensor directly gives the Burgers vector of individual dislocations with densities smeared over the reconstructed pixels. On the other hand, the tensor \(\boldsymbol{\beta}\) contains both elastic strain (stress) and lattice rotation as its symmetric and anti-symmetric parts.

The tensor fields \(\boldsymbol{\beta}\) and \(\boldsymbol{\alpha}\) can be derived from \(\boldsymbol{F}\) using equations \eqref{eq:elastic_distortion} and \eqref{eq:dislocation_density} respectively. To achieve this, however, the deformation gradient tensor field, \(\boldsymbol{F}\), must first be reconstructed using equation \eqref{eq:laue_F}. This inverse problem is nontrivial as is evident by the fact that DFXM provides 2D images of diffracted intensity, requiring a primary reconstruction of diffraction vectors, \(\boldsymbol{Q}\), from the raw data. Moreover, as discussed in section \ref{subsec:Diffraction vector measurement and acquisition geometry}, the equation system in \eqref{eq:laue_F} is underdetermined, necessitating at least three linearly independent diffraction vectors for the reconstruction of \(\boldsymbol{F}\). To address these, and similar challenges, section \ref{sec:The Dark-Field X-ray Microscope} is dedicated to outlining the critical aspects of the microscope, setting the stage for the image formation model detailed in section \ref{sec:Image formation}. Following the sampling strategy proposed in section \ref{subsec:Oblique Diffraction Geometry}, the inversion of the image formation model is then addressed in section \ref{sec:Regression scheme for determining the deformation gradient tensor field}. Finally, section \ref{sec:Numerical Example} validates the proposed framework through numerical simulations, demonstrating that the full 3D deformation gradient tensor field, \(\boldsymbol{F} = \boldsymbol{F}(\boldsymbol{x})\), can be successfully reconstructed from DFXM data.

\section{The Dark-Field X-ray Microscope}\label{sec:The Dark-Field X-ray Microscope}
\begin{figure}[H]
    \centering
    \includegraphics[width=0.87\linewidth]{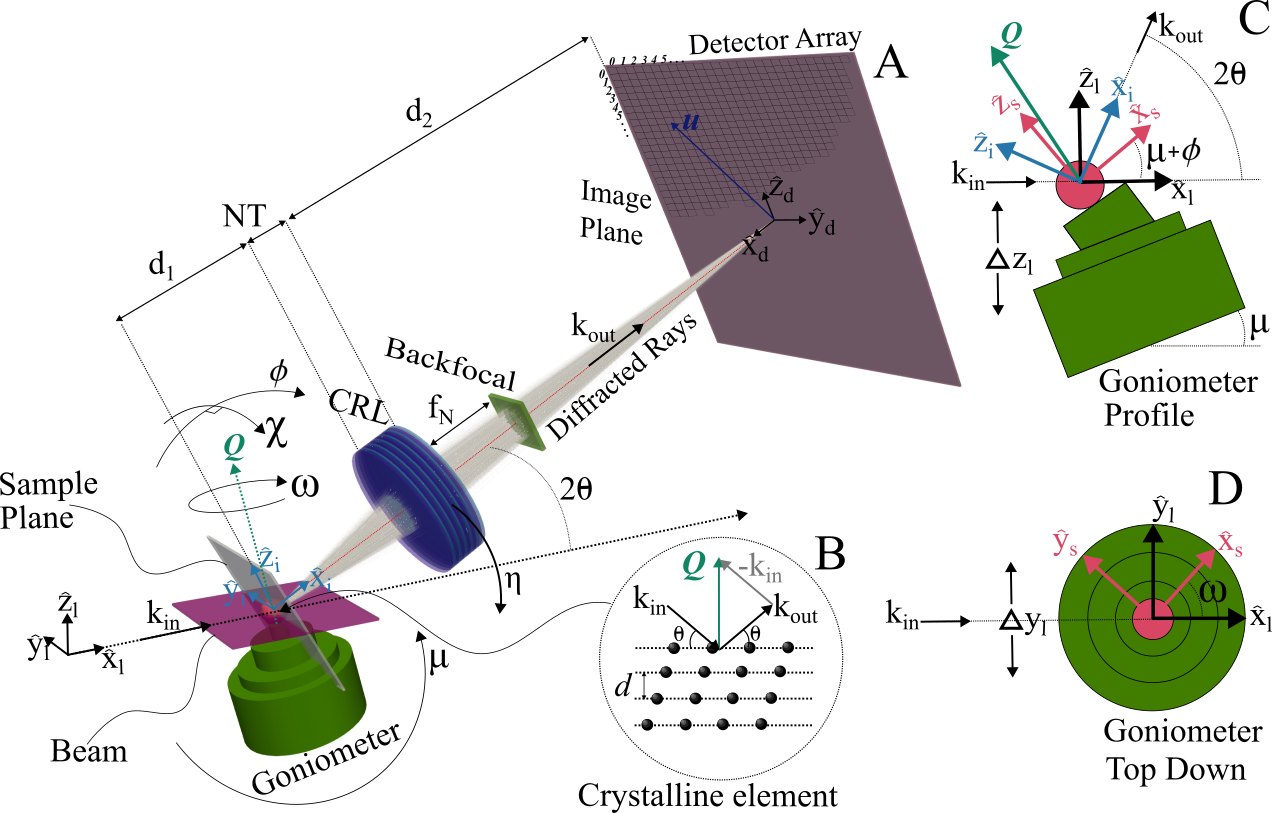}
    \caption{(A) The Dark Field X-ray microscopy geometry. A crystalline element (B) is mounted on the goniometer featuring four rotational degrees of freedom (\(\mu,\chi,\phi,\omega\)) and is illuminated by a line focused, near monochromatic, X-ray beam. By carefully adjusting the crystal orientation with respect to the incident beam, a nominal diffraction vector, \(\boldsymbol{Q}\), is brought into its diffraction condition. Placing the objective CRL at a distance \(d_1\) and using the degrees of freedom in \(\theta\) and \(\eta\) the diffracted rays pass through \(N\) lens-lets, each with thickness \(T\) and radius of curvature at apex \(R\), producing a magnified diffraction image. By adjusting the detector distance \(d_2\) the diffraction image is brought into focus. (C \& D) The sample coordinate system is illustrated to move in relation to the fixed lab frame in accordance with the goniometer rotations. The imaging system is aligned such that \(\boldsymbol{\hat{x}}_i\) is parallel to \(\boldsymbol{k'}\). When \(\omega=0\) the rotation axes corresponding to \(\phi\), \(\chi\) and \(\omega\) are mutually orthogonal. Note that depending on the setup \(\omega\) does not necessarily represent a rotation around the nominal diffraction vector \(\boldsymbol{Q}\) (as discussed in section \ref{subsec:Diffraction vector measurement and acquisition geometry}).
    }
    \label{fig:dfxm}
\end{figure}

DFXM is a synchrotron based X-ray method typically operating with a monochromatic incident beam that is condensed into a line. An X-ray objective is placed in the diffracted beam to generate a magnified view of the diffraction contrast in the illuminated sample plane. By perturbing the sample orientation and systematically recording the diffraction response on a 2D area detector placed downstream of the objective, a series of diffraction contrast images are collected. Translating the sample along the thin beam direction (perpendicular to the plane illuminated) allow for 3D space filling maps of the sample diffraction response to be collected. The diffraction contrast imaged on a single pixel of the detector in DFXM has (to a good approximation) a direct correspondence to a spatial point in the illuminated sample volume. This property of the microscope emerges form the use of an objective lens which allow all outgoing (and divergent) rays from a point in the sample plane to converge at a single point in the image plane. The optical setup can be viewed in Figure \ref{fig:dfxm}A where the four rotational degrees of freedom (\(\mu,\chi,\phi,\omega\)) of the goniometer together with the movement of the objective compound refractive lenses (CRL)  (\(\theta, \eta\)) ensures that the diffraction response from multiple crystalline domains can be reached.
By using the goniometer \(\Delta z_l\) translation degrees of freedom to step along the thin beam direction in steps equal to the beam height multiple sample sub-volumes can be stacked into a 3D volume. By utilizing the additional degree of freedom in \( \Delta y_l\) to step along the wide beam direction in steps equal to the beam width it is possible to stitch sub-volumes into large 3D maps. The corresponding series of recorded 2D diffraction images can then be thought of as an encoding of the 3D lattice deformation field. Importantly, this recorded image set corresponds to a single (and known) set of lattice planes, \(\boldsymbol{G}_{hkl}\), such that perturbations about a nominal diffraction vector \(\boldsymbol{Q}\) are probed. In section \ref{subsec:Diffraction vector measurement and acquisition geometry} we discuss how to combine several DFXM measurements to compile information from multiple reflections (i.e., from distinct \(\boldsymbol{G}_{hkl}\)).

Owing to the use of an objective the recorded diffraction images are inverted in both directions (across \(\boldsymbol{\hat{y}}_d\) and \(\boldsymbol{\hat{z}}_d\) Figure \ref{fig:dfxm}A ) and magnified by a factor \(\mathcal{M}\). Using state of the art Beryllium lens packages (as implemented at ID03 ESRF) a typical magnification values ranges from \(\mathcal{M} \approx 5\) to  \(\mathcal{M} \approx 20\). The total, effective, spatial resolution of the microscope, dependent on a series of different factors, has been found to be in the order of 100 nm \cite{Poulsen2017}. An in depth description of the microscope geometry can be found in \cite{Poulsen2017} and \cite{Poulsen2021}. For the purpose of this work we provide the relevant frames of reference in the following.

\subsection{Frames of Reference}
A fixed Cartesian laboratory coordinate system, subscripted \(l\), is introduced with \(\boldsymbol{\hat{x}}_l\) along the direct beam, \(\boldsymbol{\hat{y}}_l\) traverse to the beam, lying in the horizontal plane, and \(\boldsymbol{\hat{z}}_l\) lying in the vertical plane. The sample reference frame, subscripted \(s\), is introduced as fixed with respect to the topmost mounting point of the goniometer such that a vector in the sample frame, \(\boldsymbol{v}_s\), is transformed to the laboratory frame, \(\boldsymbol{v}_l\), by a sequence of goniometer rotations
\begin{equation}
\boldsymbol{v}_l = \boldsymbol{R_{\mu}}\boldsymbol{R}_{\omega}\boldsymbol{R}_{\chi}\boldsymbol{R}_{\phi}\boldsymbol{v}_s,
\end{equation}
where \( \boldsymbol{R_{\mu}}\in\mathbb{}R^{3\times3}\) is a positive rotation around \(\boldsymbol{\hat{y}}_l\)  of magnitude \(\mu\), \( \boldsymbol{R_{\omega}}\in\mathbb{}R^{3\times3}\) is a positive rotation around \(\boldsymbol{\hat{z}}_l\)  of magnitude \(\omega\), \( \boldsymbol{R_{\chi}}\in\mathbb{}R^{3\times3}\) is a positive rotation around \(\boldsymbol{\hat{x}}_l\)  of magnitude \(\chi\), and \( \boldsymbol{R_{\phi}}\in\mathbb{}R^{3\times3}\) is a positive rotation around \(\boldsymbol{\hat{y}}_l\) of magnitude \(\phi\) (see also Figure \ref{fig:dfxm}C-D).

Next, the crystal coordinate system is introduced, subscripted \(c\), as fixed with respect to the crystal lattice such that a vector in the crystal frame, \(\boldsymbol{v}_c\), is transformed to the sample frame, \(\boldsymbol{v}_s\), by the crystal orientation matrix \(\boldsymbol{U}\in\mathbb{R}^{3\times3}\) as
\begin{equation}
\boldsymbol{v}_s = \boldsymbol{U}\boldsymbol{v}_c.
\end{equation}
As displayed in Figure \ref{fig:dfxm}A and \ref{fig:dfxm}C the Cartesian imaging system, subscript \(i\), is defined with \(\boldsymbol{\hat{x}}_i\) along the diffracted beam such that
\begin{equation}
    \boldsymbol{v}_l = \boldsymbol{R}_{\eta}\boldsymbol{R}^T_{2 \theta}\boldsymbol{v}_i.
\end{equation}
where \( \boldsymbol{R}_{2\theta}\in\mathbb{}R^{3\times3}\) is a positive rotation around \(\boldsymbol{\hat{y}}_l\) of magnitude \(2\theta\) and \( \boldsymbol{R}_{\eta}\in\mathbb{}R^{3\times3}\) is a positive rotation around \(\boldsymbol{\hat{x}}_l\) of magnitude \(\eta\). 

The detector array holds a Cartesian coordinate system with \(\boldsymbol{\hat{y}}_d\) along the pixel columns and \(\boldsymbol{\hat{z}}_d\) along the pixel rows as seen in Figure \ref{fig:dfxm}A. The basis vector \(\boldsymbol{\hat{y}}_d\) is taken to be parallel with \(\boldsymbol{\hat{y}}_i\) while \(\boldsymbol{\hat{z}}_d\) is parallel to the projection of \(\boldsymbol{\hat{z}}_i\) unto the detector surface. The detector surface normal, \(\boldsymbol{\hat{x}}_d\), is here taken to be parallel to \(\boldsymbol{\hat{x}}_i\). For a wall mounted detector \(\boldsymbol{\hat{x}}_d\) is instead parallel to \(\boldsymbol{\hat{x}}_l\). The framework we present can operate in either of the two setups. As illustrated in Figure \ref{fig:dfxm}A points confined to the detector surface are denoted as
\begin{equation}
    \boldsymbol{u} =\begin{bmatrix}
        u\\v
    \end{bmatrix}\in \mathbb{R}^2,
    \label{eq:Pandu_def}
\end{equation}
where the coefficients \(u\) and \(v\) refer to the basis vectors \(\boldsymbol{\hat{y}}_d\) and \(\boldsymbol{\hat{z}}_d\).

In summary, the five coordinate systems used in this work are the crystal frame, sample frame, imaging frame, laboratory frame, and the detector frame.

With these definitions the explicit expressions for the incident wavevector, \(\boldsymbol{k}\), become
\begin{equation}
    \boldsymbol{k}_l = \dfrac{2\pi}{\lambda}\boldsymbol{\hat{x}}_l.
    \label{eq:k_in_def}
\end{equation}
Likewise the scattered wavevector, \(\boldsymbol{k'}\), is given as
\begin{equation}
    \boldsymbol{k'}_l = \dfrac{2\pi}{\lambda}\boldsymbol{R}_{\eta}\boldsymbol{R}^T_{2\theta}\boldsymbol{\hat{x}}_l.
    \label{eq:k_out_def}
\end{equation}

\subsection{Diffraction vector measurement and acquisition geometry}\label{subsec:Diffraction vector measurement and acquisition geometry}
It is evident that equation \eqref{eq:laue_F} is underdetermined, featuring 9 unknowns in \(\boldsymbol{F}\) but only 3 equations in \(\boldsymbol{Q}\). 
To remedy this, we propose to use a measurement scheme in which 3 or more linearly independent diffraction vectors are probed, each
featuring a distinct set of Miller indices, \(\boldsymbol{Q}_{hkl}\). We introduce the measurement matrix
\begin{equation}
    \boldsymbol{y} = \begin{bmatrix}
        \boldsymbol{Q}_{1} & \boldsymbol{Q}_{2} & \hdots & \boldsymbol{Q}_{m} 
    \end{bmatrix}\in \mathbb{R}^{3 \times m},
    \label{eq:y}
\end{equation}
together with the reference diffraction matrix
\begin{equation}
    \boldsymbol{y}^{(0)} = \begin{bmatrix}
        \boldsymbol{Q}_1^{(0)} & \boldsymbol{Q}_2^{(0)} & \hdots & \boldsymbol{Q}_m^{(0)}
    \end{bmatrix} \in \mathbb{R}^{3 \times m},
    \label{eq:y0}
\end{equation}
where
\begin{equation}
    \boldsymbol{Q}^{(0)} = 2\pi(\boldsymbol{C}^{(0)})^{-T}\boldsymbol{Q}_{hkl}.
\end{equation}
Using equations \eqref{eq:y} and \eqref{eq:y0} we find a generalized version of equation \eqref{eq:laue_F} as
\begin{equation}
      \boldsymbol{F}^{T}\boldsymbol{y} = \boldsymbol{y}^{(0)},
\end{equation}
or, alternatively, as
\begin{equation}
     \boldsymbol{y}^{(0)T} \boldsymbol{F}^{-1} = \boldsymbol{y}^T.
     \label{eq:laue_F_general}
\end{equation}
For \(m = 3\) the system in \eqref{eq:laue_F_general} has a single well defined solution if and only if \(\boldsymbol{y}^{(0)}\) is full rank (this means that the 3 selected diffraction vectors
must be linearly independent). For \(m > 3\), the system is overdetermined, which, in the presence of noise, improves the accuracy of the reconstruction, which here
is taken to be the least squares solution
\begin{equation}
    \boldsymbol{F} = \left( \left( \boldsymbol{y}^{(0)} (\boldsymbol{y}^{(0)})^T \right)^{-1} \boldsymbol{y}^{(0)} \boldsymbol{y}^T \right)^{-1}.
    \label{eq:lsq}
\end{equation}
In equation \eqref{eq:lsq} the measurements, \(\boldsymbol{y}\), represents a field of diffraction vectors, \(\boldsymbol{Q}=\boldsymbol{Q}(\boldsymbol{x})\), however, in DFXM, the primary measurement is a field of diffracted intensity, \(\boldsymbol{I}=\boldsymbol{I}(\boldsymbol{u})\). This implies that the diffraction vectors must be reconstructed from the intensity field prior to solving equation \eqref{eq:lsq}, we shall return to this topic in section \ref{sec:Q-detector}. In the here following section \ref{subsec:Oblique Diffraction Geometry}, we discuss how to reach multiple reflections in DFXM, eventually allowing us to populate \(\boldsymbol{y}\) with measurements from linearly independent diffraction vectors. In section \ref{subsec:Angular Sampling}, we instead discuss how the individual reflections must be probed, in order that we can reconstruct the diffraction vectors from the measured intensity field.

\subsection{Oblique Diffraction Geometry}\label{subsec:Oblique Diffraction Geometry}
To image a single \(\boldsymbol{Q}_{hkl}\), at a fixed X-ray wavelength, \(\lambda\), and cell matrix, \(\boldsymbol{C}^{(0)}\), the standard practice (named "simplified geometry" in \cite{Poulsen2021}) is to first bring the target \(\boldsymbol{Q}^{(0)}\) into the \(\boldsymbol{\hat{x}}_l-\boldsymbol{\hat{z}}_l\) plane using the goniometer rotation in \(\omega\). Secondly, using the base cradle tilt, \(\mu\), \(\boldsymbol{Q}^{(0)}\) is tilted to form an angle \(\pi+\theta\) to the direct beam, ensuring that the Laue equations are fulfilled while at the same time \(\eta=0\) (see also Figure \ref{fig:dfxm}B). While this procedure is standard, and allows for a wide range of \(\boldsymbol{Q}_{hkl}\) to be probed, the drawback is that each imaged reflection (different \(\boldsymbol{Q}_{hkl}\)) probes a distinct crystalline volume in the sample. Commonly, a thin layer beam is deployed in DFXM, compressed in \(\boldsymbol{\hat{z}}_l\) and extended in the horizontal \(\boldsymbol{\hat{x}}_l-\boldsymbol{\hat{z}}_l\) plane. In such a setup, 3D reconstruction (of any quantity) require compiling of space filling (3D) maps of the sample. The full dataset must then be cast to a common frame of reference by an interpolation scheme before further analysis, implying that the imaged region of interest is reduced. This can be realized by considering the intersection volume formed by a set of rotated cuboids, by necessity, the convex polyhedron that is their union must be of smaller volume than any of the ingoing cuboids. 

To lift these limitations, we use the measurement scheme described in \cite{Carsten2025}, in which the base cradle tilt is keep at \(\mu=0\) constantly while \(\omega\) is allowed to vary and \(\eta=C\) where \(C\neq0\) is constant for all recorded data. While this scheme will limit the number of available reflections, at typical X-ray energies (\(15-30\)keV), the symmetry of the imaged crystal will result in a sufficiently rich selection of (linearly independent) reflection sets. In this setting, using a layer beam, reconstruction is performed independently for each imaged \(z\)-layer, without any loss of probed sample volume. For completeness we provide, in \ref{appendix:Searching for Oblique Reflections}, the key equations used to locate a reflection set and find the corresponding \(\omega\) angles and \(\eta\) angle.

\subsection{Angular Sampling}\label{subsec:Angular Sampling}
To probe all three components of a diffraction vector, \(\boldsymbol Q\), three rotational degrees of freedom of the microscope must be used. This procedure is typically refereed to as "strain-mosa-scanning" (c.f. Section 5.2 of \cite{Poulsen2017}) and involves moving the goniometer over a fixed angular grid in \(\phi\) and \(\chi\), scanning, at each position of the goniometer, the CRL (together with the detector) over a series of angular increments in \(\theta\). Specifically, we define
\begin{equation}
   \Delta \theta =  \theta - \theta_0,
\end{equation}
where \(\theta_0\) is the nominal scattering angle (corresponding to \(\chi=\phi=0\)) and consider a regular grid of angular points in \(\Delta \theta-\phi-\chi-\)space. By recording a 2D detector image at each point on the 3D angular grid the resulting dataset of a strain-mosa-scan is 5D.

\section{Image formation in DFXM}\label{sec:Image formation}
In this section we will describe a forward model to compute 2D detector images in DFXM. The presentation corresponds to the first step in Figure \ref{fig:scheme}(a) where a discretised deformation gradient tensor field is transformed into a stack of DFXM digital images.

Similar to \cite{Poulsen2021} we adapt a geometrical optics approach to model DFXM images and utilize the same Gaussian reciprocal resolution function described in \cite{Poulsen2017}. Building on this work, we extend both the theory and implementation by removing several small angle approximations. Especially, the generalized framework that is presented allow for \(\eta\) and \(\omega \) to vary arbitrarily. This added complexity is necessary to implement the sampling scheme discussed in section \ref{subsec:Diffraction vector measurement and acquisition geometry}. Moreover, we have selected to simulate the detector point spread, the dynamic range of the camera, and the thermal and shot counting noise. The derivation of the local diffraction response of the sample is treated in section \ref{subsec:Sample diffraction response}. The propagation of the diffraction response through the optical setup to the detector plane is described in section \ref{subsec:Propagation of diffraction response to detector}. Finally, the detector point spread function and the noise model is outlined in section \ref{subsec:Detector point spread function and noise}.

\subsection{Diffraction signal strength as function of local deformation and X-ray illumination}\label{subsec:Sample diffraction response}
Operating in sample coordinates, a deformation gradient tensor field, \(\boldsymbol{F}_s=\boldsymbol{F}_s(\boldsymbol{x}_s)\) is defined over a voxelated grid and associated to a crystal reference cell matrix \(\boldsymbol{C}^{(0)}_s\). The goniometer and CRL is then calibrated to bring a target \(\boldsymbol{Q}_{hkl}\) into its diffraction condition for the nominal diffraction vector
\begin{equation}
    \boldsymbol{Q}_s^{(0)} = 2\pi\big(\boldsymbol{C}^{(0)}_s\big)^{-T}\boldsymbol{G}_{hkl}.
\end{equation}
Next, a desired perturbation of the CRL and goniometer angles is introduced and the mean of the microscope reciprocal resolution function, \(r(\boldsymbol{Q}_l) : \mathbb{R}^3 \to \mathbb{R} \), is updated accordingly. In this work we have adopted the multivariate Gaussian reciprocal resolution function first presented by \cite{Poulsen2017} and later deployed by \cite{Poulsen2021} and \cite{Borgi2024} for DFXM forward modeling using a Monte Carlo based sampling approach. Expanding on previous results we derived closed analytical expressions for the same parametric multivariate Gaussian presented by \cite{Poulsen2017}. This new result allowed us to effectively avoid the need to render the reciprocal resolution function using large sampling schemes and interpolation. The explicit analytical expression for the reciprocal resolution function, \(r(\boldsymbol{Q}_l)\), and the derivation thereof is presented in \ref{appendix:resolution}.

Introducing \(\boldsymbol{\Gamma}=\boldsymbol{R}_{\mu}\boldsymbol{R}_{\omega}\boldsymbol{R}_{\chi}\boldsymbol{R}_{\phi}\) and rearranging equation \eqref{eq:laue_F} the diffraction vector field is given in lab frame as
\begin{equation}
    \boldsymbol{Q}_l(\boldsymbol{x}_l) = \boldsymbol{\Gamma}\boldsymbol{F}_s(\boldsymbol{x}_s)^{-T}\boldsymbol{Q}_s^{(0)}, \quad \boldsymbol{x}_s = \boldsymbol{\Gamma}^T\boldsymbol{x}_l.
\end{equation}
The diffraction vector field is then mapped through the reciprocal resolution function, rendering a scalar spatial field, \(r_{hkl}(\boldsymbol{x}_l) = r(\boldsymbol{Q}_l(\boldsymbol{x}_l))\). Introducing the local photon density, \(w(\boldsymbol{x}_l): \mathbb{R}^3 \to \mathbb{R}\), the total diffraction response, \(\tau\), associated to a point, \(\boldsymbol{x}_l\), in the sample become
\begin{equation}
   \tau(\boldsymbol{x}_l) = w(\boldsymbol{x}_l) r_{hkl}(\boldsymbol{x}_l).
\end{equation}
Here \(w(\boldsymbol{x}_l)\) is assumed to vary over the sample volume to model the X-ray beam profile.

\subsection{Propagation of diffraction response to detector}\label{subsec:Propagation of diffraction response to detector}
To propagate the spatial diffraction response to the camera it will be necessary to integrate \(\tau\) over lines parallel to the diffracted rays. To simplify this line integral problem we introduce the matrix
\begin{equation}
    \boldsymbol{P} =\dfrac{-1}{\mathcal{M}}\begin{bmatrix}
        \boldsymbol{y}_d & \boldsymbol{z}_d
    \end{bmatrix}\in \mathbb{R}^{3 \times 2},
\end{equation}
which operates the CRL mapping by incorporating an image inversion and magnification though the multiplication, \(\boldsymbol{P}\boldsymbol{u}\), such that \(\boldsymbol{u}\) is mapped to a 3D coordinate in the laboratory reference frame that lies on the diffracted ray path that corresponds to the detector point \(\boldsymbol{u}\).

We are now ready to introduce the tomographic projection operator, \(\mathcal{P}: C(\mathbb{R}^3) \to C(\mathbb{R}^2)\), that maps a scalar spatial field supported in 3D real space, \(C(\mathbb{R}^3)\), to a scalar, piecewise constant, spatial field supported in 2D detector space, \(C(\mathbb{R}^2)\), as
\begin{equation}
    \mathcal{P}[\tau(\boldsymbol{x}_l)] = \int^{p_{uh}(\boldsymbol{u})}_{p_{ul}(\boldsymbol{u})}du\int^{p_{vh}(\boldsymbol{u})}_{p_{vl}(\boldsymbol{u})}dv \int^{\infty}_{-\infty}  \tau( \boldsymbol{P}\boldsymbol{u} + s_0(\boldsymbol{P}\boldsymbol{u})\boldsymbol{\hat{x}}_{i} + s\boldsymbol{\hat{x}}_{i})  ds,
    \label{eq:projection_def}
\end{equation}
were \(p_{uh}, p_{ul}, p_{vl}\) and \(p_{vh}\) are the detector pixel bounds of the pixel containing the point \(\boldsymbol{u}\) (i.e \(p_{uh}=p_{uh}(\boldsymbol{u}): \mathbb{R}^2 \to \mathbb{R}\)) and \(s_0=s_0(\boldsymbol{P}\boldsymbol{u}):\mathbb{R}^3 \to \mathbb{R}\) is the function that defines the signed distance (along \(\boldsymbol{\hat{x}}_i\)) between \(\boldsymbol{P}\boldsymbol{u}\) and the corresponding diffraction domain centroid. Specifically, a diffraction domain centroid is defined as the centroid point of the intersection between the linear path of the diffracted ray and the direct X-ray beam illumination volume. By design, this entails that the plane \(\Pi\), defined as
\begin{equation}
\Pi = \{ \boldsymbol{x}_l : \boldsymbol{x}_l = \boldsymbol{P}\boldsymbol{u} + s_0(\boldsymbol{P}\boldsymbol{u})\boldsymbol{\hat{x}}_{i}\},
\label{eq:PI}
\end{equation}
holds all diffraction domain centroids, and the integral in equation \eqref{eq:projection_def} over \(s\) is thus centered around the diffraction domain centroid.

For arbitrary values of \(\tau\) the exact projection operator \(\mathcal{P}\) can be expensive to compute. For computed tomography applications it is common to adapt a numerical approximation such that the integral over each pixel bin is approximated by the pixel centroid. We adapt this convention and use the Astra-Toolbox GPU accelerated primitives to implement \(\mathcal{P}\) \citep{Aarle2015, Aarle2016, Palenstijn2011}. Explicitly, we use the approximation
\begin{equation}
    p(\boldsymbol{u}) = \mathcal{P}[\tau(\boldsymbol{x}_l)] \approx \int^{\infty}_{-\infty} \tau( \boldsymbol{P}\boldsymbol{u} + s_0(\boldsymbol{P}\boldsymbol{u})\boldsymbol{\hat{x}}_{i} + s\boldsymbol{\hat{x}}_{i})  ds  , \quad \boldsymbol{u}\in\mathcal{D}_c,
    \label{eq:projection_def_approx}
\end{equation}
were the discretization in the approximation require that \(\boldsymbol{u}\) is taken from the set of detector pixel centroids, \(\mathcal{D}_c\), such that the projection image is piecewise constant over each pixel. Without loss of generality, the leading scale factor in \eqref{eq:projection_def_approx}, corresponding to the detector pixel area, has here been omitted for simplicity. This simplification is possible due to the later introduced exposure time factor, \(t\), which will provide an arbitrary scaling of the final image intensity values.

\subsection{Detector point spread function \& noise}\label{subsec:Detector point spread function and noise}
We introduce the detector point spread function, \(k(\boldsymbol{u}): \mathbb{R}^2 \to \mathbb{R}\), with size parameter \(\sigma_{psf}\) as
\begin{equation}
    k(\boldsymbol{u}) = c_{psf}\exp\bigg( \dfrac{-\boldsymbol{u}^T\boldsymbol{u}}{2\sigma_{psf}^2} \bigg),
\end{equation}
where \(c_{psf}\) is a normalizing constant such that \(k\) integrates to 1. The blurred diffraction image is defined as
\begin{equation}
    g(\boldsymbol{u}) = p * k
\end{equation}
where \(*\) denotes a discrete convolution, letting \(\boldsymbol{u}\) take on the detector pixel centroid values. From the blurred diffraction image a noisy diffraction image, \(f_n\), is defined as
\begin{equation}
    g_n(\boldsymbol{u}) = t g(\boldsymbol{u}) + n(g(\boldsymbol{u}), \sigma^2_{thermal}, \mu_{thermal}),
\end{equation}
where the function \(n\) represents a combination of Poisson shot noise, with signal-to-noise ratio \(SNR=\sqrt{g}\), and Gaussian thermal noise, with mean \(\mu_{thermal}\) and variance \(\sigma^2_{thermal}\). Here, \(t\) is proportional to the exposure time of the camera and regulates the overall signal-to-noise ratio of the simulated data. Finally, the dynamic range of the camera is simulated by binning \(g_n\) to unsigned 16-bit integers. The resulting digital image is 
\begin{equation}
    I(\boldsymbol{u}) = \text{round}_{16}[ g_n(\boldsymbol{u}) ], \quad \boldsymbol{u}\in\mathcal{D}_c.
\end{equation}

\section{Regression scheme for reconstructing the deformation gradient tensor field}\label{sec:Regression scheme for determining the deformation gradient tensor field}
In this section we show how a collection of DFXM images can be analyzed to reconstruct the deformation gradient tensor field. The exposition corresponds to each the four steps in Figure \ref{fig:scheme}(b)-(e). The proposed regression scheme is structured as a sequential transformation; first, the mean diffraction angles in \(\phi, \chi\) and \( \Delta\theta\) are extracted from the photon count intensities recorded on the detector (b). Second, for each probed reflection \((hkl)\), the mean diffraction angles in \(\phi, \chi\) and \( \Delta\theta\) are transformed into diffraction vectors in detector space, such that each pixel corresponds to a single volume-averaged diffraction vector \(\boldsymbol{Q}\) (c). Third, the diffraction vectors are back-propagated through the CRL lens into sample space, incorporating the \(\omega\)-rotations of the various \((hkl)\) reflections (d). This procedure assigns a set of three or more diffraction vectors to each voxel within the sample volume. Finally, these voxel-specific diffraction vector sets are transformed into deformation gradient tensors using equation \eqref{eq:lsq} (e). We will discuss each of these topics in order, starting in section \ref{sec:CoM} with the extraction of mean diffraction angles in \(\phi, \chi\) and \( \Delta\theta\) (i.e Figure \ref{fig:scheme}b).

\subsection{Extracting the mean diffraction angles in \(\phi, \chi\) and \( \Delta\theta\) positions}\label{sec:CoM}
\begin{figure}[H]
    \centering
    \includegraphics[width=0.8\linewidth]{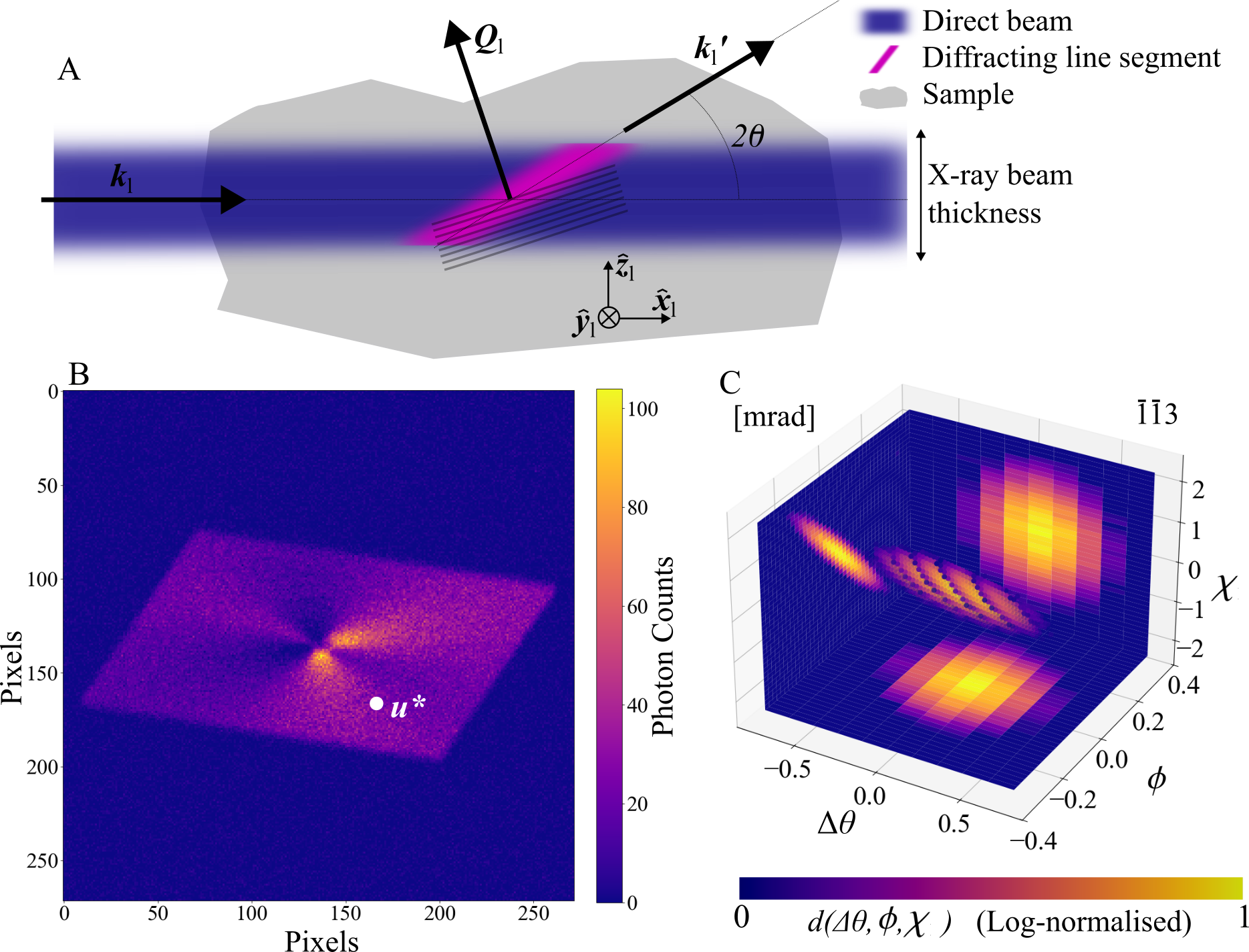}
    \caption{Diffraction contrast in DFXM. (A) Schematic of intersection between direct X-ray (blue) and sample (gray) during diffraction. The cerise subdomain diffracts along \(\boldsymbol{k'}\) and strikes the detector in sub-figure (B) at \(\boldsymbol{u}^*\).  
    (B) Simulated detector image at \(\Delta \theta=\)-0.42 mrad,
    \(\phi=\)0.46 mrad, \(\chi=\)0.067 mrad for the single crystal phantom described in section \ref{sec:Phantom} featuring a single, centered, straight edge dislocation line. The angular distribution of scattered intensity, illustrated to originate from a diffracting line segment in the sample in (A), is displayed in (C) and corresponds to the detector pixel marked as \(\boldsymbol{u}^*\) in (B). The sample was scanned over a discrete (and non-uniform) grid in \(\Delta \theta\),
    \(\phi\), \(\chi\), and the resulting angular distribution of scattered intensity in figure (C) is seen to be point-wise supported, with the coarser angular step-size along the \(\Delta \theta\) direction. For visual purposes all points above a fixed threshold in the scatter plot of (C) have been plotted together with the three Cartesian projections of the point could illustrated as 2D intensity profiles. }
    \label{fig:angmap}
\end{figure}

Given a sequence of DFXM images, \(I_{1}(\boldsymbol{u}), I_{2}(\boldsymbol{u}), .., I_{k}(\boldsymbol{u})\), each associated to an angular goniometer and CRL setting we define the 5D data function, \(D(\boldsymbol{u}, \Delta \theta, \phi, \chi): \mathbb{R}^5 \to \mathbb{R}\), to contain the full dataset of photon count values (for a single reflection). For a fixed coordinate on the detector, \(\boldsymbol{u}=\boldsymbol{u}^*\), the resulting function \(d(\Delta \theta, \phi, \chi)=D(\boldsymbol{u}^*, \Delta \theta, \phi, \chi): \mathbb{R}^3 \to \mathbb{R}\) describes the angular distribution of scattered intensity originating from a diffracting line segment in the sample. Specifically, \(d(\Delta \theta, \phi, \chi)\) is associated to a line segment in the sample, defined as the clip between a ray parallel to \(\boldsymbol{k'}\) and the direct X-ray beam domain as illustrated in Figure \ref{fig:angmap}A. With the present embodiment of the instrument at ID03 the direct space resolution element is asymmetric with the long direction being defined by the vertical beam-height. 

In the following analysis we aim to reconstruct diffraction vectors, \(\boldsymbol{Q}\), based on the diffracted intensity distributions, \(d(\Delta \theta, \phi, \chi)\). We again stress that the procedure outlined here operates under the approximation of kinematic scattering and the influence of dynamical scattering on this analysis is out of scope of this work (see also the work of \cite{Borgi2024} for a comparison of the kinematic and dynamic diffraction approximations).

Collecting the angular degrees of freedom in \(\phi\), \(\chi\) and \(\Delta \theta\) in a single variable as
\begin{equation}
    \boldsymbol{\psi} = \begin{bmatrix}
        \Delta \theta \\ \phi \\ \chi
    \end{bmatrix} \in \mathbb{R}^3,
\end{equation}
we may compactly denote \(D(\boldsymbol{u}, \Delta \theta, \phi, \chi) = D(\boldsymbol{u}, \boldsymbol{\psi})\). Next, we introduce the probability density function 
\begin{equation}
    p_{\Psi}(\boldsymbol{\psi}| \boldsymbol{u}) = \dfrac{1}{c_{\Psi}}D(\boldsymbol{u}, \boldsymbol{\psi}),
\end{equation}
where \(c_{\Psi}\) is a normalizing constant such that
\begin{equation}
    \int_{\forall \boldsymbol{\psi}} p_{\Psi}(\boldsymbol{\psi}| \boldsymbol{u}) d\boldsymbol{\psi} = 1
\end{equation}
The spatial map, \(\boldsymbol{\Bar{d}}_{\psi}(\boldsymbol{u}): \mathbb{R}^2 \to \mathbb{R}^3\), mapping a detector point, \(\boldsymbol{u}\), to a set of mean diffraction angles, \(\boldsymbol{\psi}\), can now be introduced as the expectation value
\begin{equation}
    \boldsymbol{\Bar{d}}_{\psi}(\boldsymbol{u}) = \mathbb{E}[\boldsymbol{\Psi}|\boldsymbol{u}] = \int_{\forall \boldsymbol{\psi}} \boldsymbol{\psi}p_{\Psi}(\boldsymbol{\psi}| \boldsymbol{u}) d\boldsymbol{\psi}
    \label{eq:CoM}
\end{equation}
where \(\mathbb{E}\) is the expectation operator and \(\boldsymbol{\Psi}\) is the stochastic variable associated to \(p_{\Psi}\) such that \(\boldsymbol{\psi}\) is a realization of \(\boldsymbol{\Psi}\).

The map \(\boldsymbol{\Bar{d}}_{\psi}(\boldsymbol{u})\) in equation \eqref{eq:CoM} constitutes the primary quantity extracted in our analysis. It describes the mean diffraction angles in \(\phi, \chi\) and \( \Delta\theta\) over the detector surface and serve as the basis for the reconstruction of diffraction vectors, \(\boldsymbol{Q}\). Before proceeding to do so, however, we stress that the value of \(\boldsymbol{\Bar{d}}_{\psi}(\boldsymbol{u})\) is the weighted average of the diffraction angles in \(\phi, \chi\) and \( \Delta\theta\) approximately collected on the spatial domain illustrated in Figure \ref{fig:angmap}A. The weighting is here a property of the beam intensity profile distribution as we shall return to in section \ref{sec:Backpropagation}. This means that any proceeding reconstructions must be thought of as averaged quantities over subdomains in the sample volume. Only in the limit, when the thickness of the beam approaches zero, does \(d_{com}(\boldsymbol{u})\) approach a point-wise measurement of the mean diffraction angles in \(\phi, \chi\) and \( \Delta\theta\).

We stress that for large angular perturbations in \(\phi\) and or \(\chi\) the assumption that the angular intensity distribution, \(D(\boldsymbol{u}, \boldsymbol{\psi})\), is associated to a connected line segment in the sample breaks down. Similarly, for a large sample radii this linear-path approximation will deteriorate with increasing distance from the optical axis. However, for moderate angular perturbations in \(\phi\) and or \(\chi\) and a limited sample region of interest, centered on the optical axis, our approximation will result in a moderate spatial blurring of the diffraction contrast. In this work we probe the end-to-end effect of this approximation by including in our forward model the effect of the perturbations in \(\phi\) and \(\chi\) on the diffraction response. This means that every modeled diffraction image is associated to a slightly different sample tilt in \(\phi\) and \(\chi\). It is worth noting that the validity of this approximation - that the diffraction response of a single pixel on the detector is associated to a connected line segment in the sample - is in fact relevant for single reflection DFXM imaging applications such as local orientation mapping by \(\phi-\chi\)-scanning.

\subsection{Reconstructing Diffraction Vectors, \(\boldsymbol{Q}\)}\label{sec:Q-detector}
With the mean diffraction angles in \(\phi, \chi\) and \( \Delta\theta\) known we now move on in our analysis to the step corresponding to Figure \ref{fig:scheme}(c). The task is now to convert \(\boldsymbol{\Bar{d}}_{\psi}(\boldsymbol{u})\) into diffraction vectors, \(\boldsymbol{Q}\). To this end we shall start by assuming the existence of an injective intermediate map, \(\boldsymbol{f}(\boldsymbol{\psi}): \mathbb{R}^3 \to \mathbb{R}^3\), such that
\begin{equation}
    \boldsymbol{f}(\boldsymbol{\psi}) = \boldsymbol{Q}_s,
    \label{eq:f_def}
\end{equation}
where \(\boldsymbol{Q}_s\) is a diffraction vector in sample coordinates. By definition it must hold that
\begin{equation}
p_{\Psi}(\boldsymbol{\psi}| \boldsymbol{u}) = p_Q( \boldsymbol{f}(\boldsymbol{\psi})| \boldsymbol{u} ) = p_Q( \boldsymbol{Q}_s | \boldsymbol{u} ),
\label{eq:p_L=p_Q}
\end{equation}
for some (unknown) probability density function \(p_Q\). The equality of equation \eqref{eq:p_L=p_Q} together with the definition in \eqref{eq:f_def} implies that
\begin{equation}
    \mathbb{E}[\boldsymbol{Q}_s| \boldsymbol{u}] = \int_{\forall \boldsymbol{Q}_s} \boldsymbol{Q}_s p_{\boldsymbol{Q}}(\boldsymbol{Q}_s| \boldsymbol{u}) d\boldsymbol{Q}_s = \int_{\forall \boldsymbol{\psi}}  \boldsymbol{f}(\boldsymbol{\psi}) p_{\Psi}(\boldsymbol{\psi}| \boldsymbol{u}) d\boldsymbol{\psi},
    \label{eq:EQ_general}
\end{equation}
where \(\mathbb{E}\) is the expectation operator. When the map, \(\boldsymbol{f}\), is known, it is clear from equation \eqref{eq:EQ_general} that the computation of \(\mathbb{E}[\boldsymbol{Q}_s| \boldsymbol{u}]\) can be achieved by mapping all angular points, \(\boldsymbol{\psi}\), on the support of \(D(\boldsymbol{u}, \boldsymbol{\psi})\), to diffraction vectors, \(\boldsymbol{Q}_s=\boldsymbol{f}(\boldsymbol{\psi})\), and performing the right hand side integral in \eqref{eq:EQ_general}. To determine \(\boldsymbol{f}\) we must consider the Laue equations \eqref{eq:laue}. Combining equations \eqref{eq:elastic_diffraction_vector}, \eqref{eq:k_in_def} and \eqref{eq:k_out_def}, for a fixed \(\theta\), it follows that
\begin{equation}
    \boldsymbol{Q}_l = \dfrac{2\pi}{\lambda}\big(\boldsymbol{R}_{\eta}\boldsymbol{R}^T_{2\theta} - \boldsymbol{I})\boldsymbol{\hat{x}}_l,
\end{equation}
where \(\lambda\) is the mean wavelength. Considering that the sample space diffraction vector must satisfy
\begin{equation}
    \boldsymbol{Q}_l = \boldsymbol{\Gamma}\boldsymbol{Q}_s = \boldsymbol{\Gamma}\boldsymbol{f}(\boldsymbol{\psi}),
\end{equation}
we find the explicit map
\begin{equation}
    \boldsymbol{f}(\boldsymbol{\psi}) =  \dfrac{2\pi}{\lambda}\boldsymbol{R}^T_{\phi}\boldsymbol{R}^T_{\chi} \boldsymbol{R}^T_{\omega}\boldsymbol{R}^T_{\mu}\big(\boldsymbol{R}_{\eta}\boldsymbol{R}^T_{2\theta} - \boldsymbol{I})\boldsymbol{\hat{x}}.
    \label{eq:f}
\end{equation}
Equations \eqref{eq:f} and \eqref{eq:EQ_general} provide a way to compute \(\mathbb{E}[\boldsymbol{Q}_s| \boldsymbol{u}]\) for arbitrary deformation magnitudes and can be seen as a generalization of \cite{Carsten2025} in which the diffraction vector is approximated by first order Taylor expansion as
\begin{equation}
    \boldsymbol{Q}_s = \boldsymbol{f}(\boldsymbol{\psi}) \approx \boldsymbol{f}(\boldsymbol{0}) + \bigg(\dfrac{\partial \boldsymbol{f}}{\partial \boldsymbol{\psi}}\bigg\rvert_{\boldsymbol{\psi}=\boldsymbol{0}}\bigg)\boldsymbol{\psi}.
    \label{eq:taylor_f}
\end{equation}
Inserting \eqref{eq:taylor_f} in \eqref{eq:EQ_general} we find that
\begin{equation}
    \mathbb{E}[\boldsymbol{Q}_s| \boldsymbol{u}] \approx \boldsymbol{f}(\boldsymbol{0}) + \bigg(\dfrac{\partial \boldsymbol{f}}{\partial \boldsymbol{\psi}}\bigg\rvert_{\boldsymbol{\psi}=\boldsymbol{0}} \bigg)\int_{\forall \boldsymbol{\psi}} \boldsymbol{\psi}p_{\Psi}(\boldsymbol{\psi}| \boldsymbol{u}) d\boldsymbol{\psi} = \boldsymbol{f}(\boldsymbol{0}) + \bigg(\dfrac{\partial \boldsymbol{f}}{\partial \boldsymbol{\psi}}\bigg\rvert_{\boldsymbol{\psi}=\boldsymbol{0}}\bigg)\mathbb{E}[\boldsymbol{\Psi}|\boldsymbol{u}]
    \label{eq:EQ_taylor}.
\end{equation}
By identifying that the right hand side of \eqref{eq:EQ_taylor} is the Taylor expansion of \(\boldsymbol{f}(\mathbb{E}[\boldsymbol{\Psi}|\boldsymbol{u}])\) it follows that the approximation given by \cite{Carsten2025} can be practically realized as
\begin{equation}
     \mathbb{E}[\boldsymbol{Q}_s| \boldsymbol{u}] \approx \boldsymbol{f}(\mathbb{E}[\boldsymbol{\Psi}|\boldsymbol{u}]) = \boldsymbol{f}(\boldsymbol{\Bar{d}}_{\psi}(\boldsymbol{u})).
    \label{eq:f(E[L])=E[Q_s]}
\end{equation}
 In contrast to the work of \cite{Carsten2025}, by directly applying \(\boldsymbol{f}\) to \(\boldsymbol{\Bar{d}}_{\psi}(\boldsymbol{u})\) the formulation proposed in equation \eqref{eq:f(E[L])=E[Q_s]} effectively avoids computing any partial derivatives. In this work we have opted to used equation \eqref{eq:f(E[L])=E[Q_s]} to compute \(\mathbb{E}[\boldsymbol{Q}_s| \boldsymbol{u}]\), which represents an excellent approximation for \(\boldsymbol{\psi}<<1\). In \ref{appendix:Taylor expansion of f} we provide numerical estimates of the error introduced by this approximation for increasing \(||\boldsymbol{\psi}||_2\). We stress that while we use an approximate form of \eqref{eq:EQ_general} for the results presented in this paper, the map we have provided in equation \eqref{eq:f} allow in general for \eqref{eq:EQ_general} to be evaluated without approximation. For highly deformed samples, when the scanned range in \(\boldsymbol{\psi}\) is large, this distinction become important.

 As previously discussed, the diffraction vector field, \(\mathbb{E}[\boldsymbol{Q}_s| \boldsymbol{u}]\), obtained over the detector space represents an average property over the associated line segments in the sample (weighted by the local photon density). This follows from the fact that \(p_Q\) and \(p_{\Psi}\) are probability density functions, making equation \eqref{eq:EQ_general} a direct application of the expectation value definition, as denoted by \(\mathbb{E}[\cdot]\). This property is critical for the next step of our analysis, which involves the backpropagation of \(\mathbb{E}[\boldsymbol{Q}_s | \boldsymbol{u}]\) from detector space to sample space. With the tomographic approximations made in section \ref{subsec:Propagation of diffraction response to detector} we may write
\begin{equation}
    \mathbb{E}[\boldsymbol{Q}_s | \boldsymbol{u}] = \dfrac{\mathcal{P}[w(\boldsymbol{x}_l)\boldsymbol{\Gamma}^T\boldsymbol{Q}_l(\boldsymbol{x}_l)]}{\mathcal{P}[w(\boldsymbol{x}_l)]}.
    \label{eq:E[Q]_projected}
\end{equation}
Equation \eqref{eq:E[Q]_projected} provides the central link between the sample space diffraction vector field, \(\boldsymbol{Q}_s\), and the detector space field, \(\mathbb{E}[\boldsymbol{Q}_s | \boldsymbol{u}]\).

\subsection{Backpropagation from detector to sample diffraction vectors}\label{sec:Backpropagation}
We have now derived the per-detector-pixel mean diffraction vectors, \(\mathbb{E}[\boldsymbol{Q}_s| \boldsymbol{u}]\), by analyzing the detector intensity as a function of \(\chi, \phi\), and \(\Delta \theta\). The goal of our analysis, however, is to obtain a spatial deformation field in the sample domain rather than in detector space. It is therefore crucial to relate the detector pixel mean diffraction vectors, \(\mathbb{E}[\boldsymbol{Q}_s| \boldsymbol{u}]\), to diffraction vectors in the sample or lab space, \(\boldsymbol{Q}(\boldsymbol{x})\). The existence and uniqueness of such a link between \(\mathbb{E}[\boldsymbol{Q}_s| \boldsymbol{u}]\) and \(\boldsymbol{Q}(\boldsymbol{x})\) are nontrivial to establish due to the finite X-ray beam thickness along \(\boldsymbol{\hat{z}}_l\) in DFXM. 

As we shall see, the mapping between detector space diffraction vectors and sample space diffraction vectors cannot be uniquely established for arbitrary deformations. However, when the deformation field varies approximately linearly across the X-ray beam thickness, \(\boldsymbol{Q}(\boldsymbol{x})\) can be recovered from \(\mathbb{E}[\boldsymbol{Q}_s| \boldsymbol{u}]\). The conclusion from this analysis is that, in DFXM experiments, the beam thickness along \(\boldsymbol{\hat{z}}_l\) should be sufficiently small to ensure that the imaged deformation gradient tensor field, \(\boldsymbol{F}(\boldsymbol{x})\), varies weakly over the diffracting line segments in the sample. Notably, variations of \(\boldsymbol{F}(\boldsymbol{x})\) in the transverse directions, within the beam plane, are not subject to this restriction. 

Furthermore, when this condition is not met and \(\boldsymbol{F}(\boldsymbol{x})\) varies strongly over the diffracting domain, an error is introduced into \(\boldsymbol{Q}(\boldsymbol{x})\), which will propagate into the reconstructed deformation gradient tensor. The magnitude of this error depends on the variation of \(\boldsymbol{F}(\boldsymbol{x})\) across the diffracting domain, with larger errors expected for increased beam widths (in \(\boldsymbol{\hat{z}}_l\)) and higher curvatures in \(\boldsymbol{F}(\boldsymbol{x})\) (i.e., larger second-order spatial derivatives).

\subsubsection{Derivation of a map between \(\mathbb{E}[\boldsymbol{Q}_s| \boldsymbol{u}]\) and  \(\boldsymbol{Q}(\boldsymbol{x})\)}\label{sec:Backpropagation Derivation of a map between}
In the following we shall prove that when the beam profile is symmetric across the diffracting line segments in the sample, and, additionally, the deformation field variation is odd with respect to the centroid plane an exact reconstruction of \(\boldsymbol{Q}(\boldsymbol{x})\) can be achieved across the diffraction domain centroid plane,\(\Pi\) (see equation \eqref{eq:PI}). To this end, let
\begin{equation}
    \boldsymbol{\Gamma}^T\boldsymbol{Q}_l(\boldsymbol{x}_l) \approx \boldsymbol{c}_0( \boldsymbol{x}_l ) + \boldsymbol{\kappa}( \boldsymbol{x}_l ),
    \label{eq:Q_odd}
\end{equation}
over the X-ray beam support where \( \boldsymbol{\kappa}: \mathbb{R}^3 \to \mathbb{R}^3\) is odd across the plane \(\Pi\), and \(\boldsymbol{c}_0: \mathbb{R}^3 \to \mathbb{R}^3\) is constant over each diffracting line segment. Using the notation accompanying equation \eqref{eq:projection_def_approx} we may parameterize the spatial coordinate as \(\boldsymbol{x}_l=\boldsymbol{x}_l(\boldsymbol{u}, s) = \boldsymbol{P}\boldsymbol{u} + s_0\boldsymbol{\hat{x}}_i + s\boldsymbol{\hat{x}}_i\) and formally define
\begin{equation}
    \dfrac{\partial \boldsymbol{c}_0( \boldsymbol{x}_l )}{\partial s} = \dfrac{\partial \boldsymbol{c}_0( \boldsymbol{x}_l )}{\partial \boldsymbol{x}_l}\dfrac{\partial \boldsymbol{x}_l}{\partial s} = \dfrac{\partial \boldsymbol{c}_0( \boldsymbol{x}_l )}{\partial \boldsymbol{x}_l}\boldsymbol{\hat{x}}_i = \boldsymbol{0},
    \label{eq:grad_c0_is_zero}
\end{equation}
implying that the gradient of \(\boldsymbol{c}_0\) is orthogonal to \(\boldsymbol{\hat{x}}_i\). The requirement on \( \boldsymbol{\kappa}\) can now be formalized as
\begin{equation}
     \boldsymbol{\kappa}( \boldsymbol{P}\boldsymbol{u} + s_0\boldsymbol{\hat{x}}_i + s\boldsymbol{\hat{x}}_i ) = - \boldsymbol{\kappa}( \boldsymbol{P}\boldsymbol{u} + s_0\boldsymbol{\hat{x}}_i - s\boldsymbol{\hat{x}}_i ), \quad \forall s.
    \label{eq:kappa_odd}
\end{equation}
Equation \eqref{eq:Q_odd} represent a wide family of diffraction vector fields, including, but not limited to linear variation over the diffraction domain (along \(s\)) together with arbitrarily non-linear variations in the traverse plane, \(\Pi\). Moreover, the approximation that is equation \eqref{eq:Q_odd} is to be interpreted as a local approximation that only need to be valid over the diffraction domains. 

We proceed by inserting equation \eqref{eq:Q_odd} into equation \eqref{eq:E[Q]_projected} to find
\begin{equation}
    \mathbb{E}[\boldsymbol{Q}_s | \boldsymbol{u}] = \dfrac{\mathcal{P}[w(\boldsymbol{x}_l)\boldsymbol{c}_0( \boldsymbol{x}_l )]}{\mathcal{P}[w(\boldsymbol{x}_l)]} +
 \dfrac{\mathcal{P}[w(\boldsymbol{x}_l) \boldsymbol{\kappa}( \boldsymbol{x}_l )]}{\mathcal{P}[w(\boldsymbol{x}_l)]},
    \label{eq:backprop_1}
\end{equation}
were the linearity of \(\mathcal{P}\) was used. For a typical X-ray beam profile, featuring a Gaussian intensity decay along \(\boldsymbol{\hat{z}}_l\), the additional symmetry
\begin{equation}
    w( \boldsymbol{P}\boldsymbol{u} + s_0\boldsymbol{\hat{x}}_i + s\boldsymbol{\hat{x}}_i ) = w( \boldsymbol{P}\boldsymbol{u} + s_0\boldsymbol{\hat{x}}_i - s\boldsymbol{\hat{x}}_i ), \quad \forall s,
\end{equation}
renders \(w( \boldsymbol{x}_l )\boldsymbol{\kappa}( \boldsymbol{x}_l )\) odd across \(\Pi\). Considering the definition of \(\mathcal{P}\) in equation \eqref{eq:projection_def} it therefore follows that
\begin{equation}
    \mathcal{P}[w(\boldsymbol{x}_l) \boldsymbol{\kappa}( \boldsymbol{x}_l )] = 0.
\end{equation}
Equation \eqref{eq:backprop_1} now reduces to
\begin{equation}
    \mathbb{E}[\boldsymbol{Q}_s | \boldsymbol{u}] = \dfrac{\mathcal{P}[w(\boldsymbol{x}_l)\boldsymbol{c}_0( \boldsymbol{x}_l )]}{\mathcal{P}[w(\boldsymbol{x}_l)]}.
    \label{eq:backprop_2}
\end{equation}
Considering next that \(\boldsymbol{c}_0\) is constant over each diffracting line segment we find from the definition in equation \eqref{eq:projection_def} that
\begin{equation}
    \mathbb{E}[\boldsymbol{Q}_s | \boldsymbol{u}] = \dfrac{\boldsymbol{c}_0(\boldsymbol{P}\boldsymbol{u} + s_0\boldsymbol{\hat{x}}_i)\int^{\infty}_{-\infty}w(\boldsymbol{P}\boldsymbol{u} + s_0\boldsymbol{\hat{x}}_i - s\boldsymbol{\hat{x}}_i) ds}{\int^{\infty}_{-\infty}w(\boldsymbol{P}\boldsymbol{u} + s_0\boldsymbol{\hat{x}}_i - s\boldsymbol{\hat{x}}_i) ds} = \boldsymbol{c}_0(\boldsymbol{P}\boldsymbol{u} + s_0\boldsymbol{\hat{x}}_i)
    \label{eq:EQ_eq_c0}
\end{equation}
Dropping the parametrisation of \(\boldsymbol{x}_l=\boldsymbol{x}_l(\boldsymbol{u},s)\) and considering that \(\boldsymbol{\kappa}\) is odd it follows from equation \eqref{eq:Q_odd} and \eqref{eq:EQ_eq_c0} that
\begin{equation}
    \boldsymbol{\Gamma}^T\boldsymbol{Q}_l(\boldsymbol{x}_l) = \mathbb{E}[\boldsymbol{Q}_s | \boldsymbol{P}^\dagger\boldsymbol{x}_l], \quad \forall \boldsymbol{x}_l \in \Pi
    \label{eq:recon_Q}
\end{equation}
where \(\boldsymbol{P}^\dagger\) is the pseudo-inverse of \(\boldsymbol{P}\) such that \(\boldsymbol{P}^\dagger\boldsymbol{P}=\boldsymbol{I}\in\mathbb{R}^{2\times2}\). Equation \eqref{eq:recon_Q} is the sought backpropagation of \(\mathbb{E}[\boldsymbol{Q}_s | \boldsymbol{u}]\) to the centroid plane of diffraction, \(\Pi\). In conclusion, we have now proven, for a general class of deformations, that the diffraction vector field, \(\mathbb{E}[\boldsymbol{Q}_s | \boldsymbol{u}]\), existing on the detector space, can be directly associated to the real space diffraction vector field of the sample.

Finally we discuss in section \ref{sec:Local LSQ} how to unify the result in equation \eqref{eq:recon_Q} to a setting featuring multiple reflections, allowing us to give the least squares reconstruction of \(\boldsymbol{F}_s(\boldsymbol{x}_s)\), which is the goal of our regression scheme. We also note that while we have here formulated both the image formation scheme and the backpropagation scheme in terms of the sample rotating with the goniometer, for computational purposes, and without loss of generality, we have implemented the reverse motion, allowing the X-ray beam and detector to orbit the sample.

\subsection{Local Least Squares Squares Reconstruction of \(\boldsymbol{F}_s\)}\label{sec:Local LSQ}
With the spatial diffraction vector field, \(\boldsymbol{Q}_s(\boldsymbol{x}_s)\), now available we are ready to reconstruct the deformation gradient tensor field, \(\boldsymbol{F}_s(\boldsymbol{x}_s)\). This final part of our analysis corresponds to Figure \ref{fig:scheme}(e).

Consider a setting were multiple reflections have been recorded, each at a different goniometer setting, we may unify the result in equation \eqref{eq:recon_Q} to the same sample space via the rotation
\begin{equation}
    \boldsymbol{Q}_s(\boldsymbol{x}_s) =  \boldsymbol{\Gamma}^T\boldsymbol{Q}_l(\boldsymbol{\Gamma}^T\boldsymbol{x}_l) = \mathbb{E}[\boldsymbol{Q}_s | \boldsymbol{P}^\dagger\boldsymbol{\Gamma}^T\boldsymbol{x}_l], \quad \forall \boldsymbol{x}_l \in \Pi,
    \label{eq:Qs(xs)}
\end{equation}
where \(\boldsymbol{\Gamma}\) is approximated by the nominal settings of the goniometer for the considered reflection, i.e \(\phi=\chi=\Delta \theta=0\), \(\omega\neq0\).

Considering \(m\) distinct reflections the measurement matrix
\begin{equation}
    \boldsymbol{y}_s = \boldsymbol{y}_s(\boldsymbol{x}_s) = \begin{bmatrix}
        \boldsymbol{Q}^{(0)}_s(\boldsymbol{x}_s) & \boldsymbol{Q}^{(1)}_s(\boldsymbol{x}_s) & \hdots & \boldsymbol{Q}^{(m)}_s(\boldsymbol{x}_s),
    \end{bmatrix}
\end{equation}
can now be produced in accordance with equation \eqref{eq:y} (note that \(\boldsymbol{\Gamma}^{(0)}, \boldsymbol{\Gamma}^{(1)},...\) varies between reflections). Here the notation \(\boldsymbol{y}_s=\boldsymbol{y}_s(\boldsymbol{x}_s)\) implies that diffraction vector measurement matrix is changing across the sample domain. We define the constant reference diffraction matrix, \(\boldsymbol{y}_s^{(0)}\), using equation \eqref{eq:y0}, and apply the least squares operator in equation \eqref{eq:lsq} as the point wise operation
\begin{equation}
    \boldsymbol{F}_s(\boldsymbol{x}_s) = \left( \left( \boldsymbol{y}_s^{(0)} (\boldsymbol{y}_s^{(0)})^T \right)^{-1} \boldsymbol{y}_s^{(0)} \boldsymbol{y}_s^T \right)^{-1}.
\end{equation}
This concludes our regression scheme. We note that from the reconstructed deformation gradient tensor field, \(\boldsymbol{F}_s(\boldsymbol{x}_s)\), the elastic distortions can be trivially recovered as \(\boldsymbol{\beta}_s(\boldsymbol{x}_s) = \boldsymbol{F}_s(\boldsymbol{x}_s) -\boldsymbol{I}\), likewise, the dislocation density tensor field, \(\boldsymbol{\alpha}_s(\boldsymbol{x}_s)\), can be computed using equation \eqref{eq:dislocation_density}.

\section{Numerical validation example}\label{sec:Numerical Example}
We have now described both the framework used to generate synthetic DFXM diffraction images as well as the regression scheme used to analyze such data. The 5 steps in Figure \ref{fig:scheme}(a)-(e) are thus complete, and we proceed by dedicating the following section to validating our proposed models using a numerical example. To this end the forward model described in section \ref{sec:Image formation} was used to generate synthetic noisy DFXM diffraction images. The deformation gradient tensor field for the simulation corresponds to a single straight edge dislocation, as detailed in section \ref{sec:Phantom}. The microscope setup, including the CRL, detector, and goniometer parameters, is provided in section \ref{sec:Setup}. Using the regression framework outlined in section \ref{sec:Regression scheme for determining the deformation gradient tensor field}, we reconstructed the deformation gradient tensor field over three consecutive \(z\)-slices of the sample. In section \ref{sec:Reconstructions}, we compare the ground truth deformation to the reconstructed deformation for a slice in the sample volume and give the residual error.

\subsection{Deformation gradient tensor field phantom}\label{sec:Phantom}
\begin{figure}[H]
    \centering
    \includegraphics[width=0.95\linewidth]{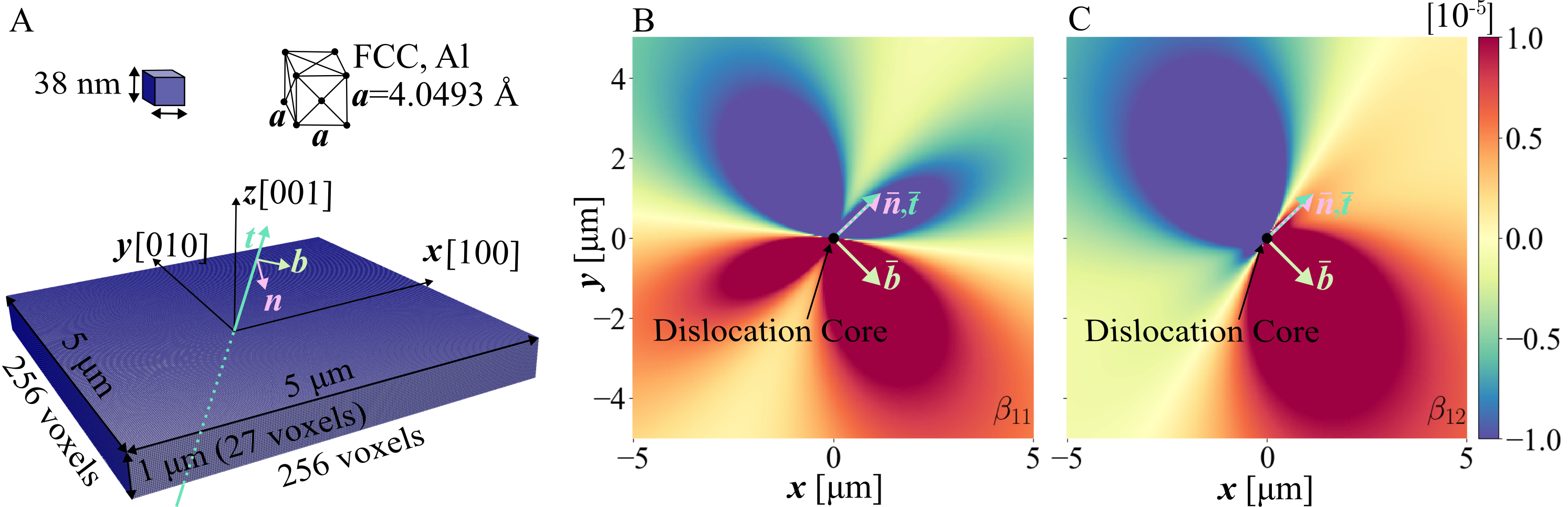}
    \caption{Phantom used for numerical validation: A 25 \(\mu\)m\(^3\) single crystal FCC Aluminum sample was meshed with a voxel size of \(\sim\)38 nm (A). The cubic crystal unit cell was aligned with the Cartesian lab frame and a straight edge dislocation was introduced to pierce the origin, with line direction \( \boldsymbol{t} \), Burgers vector \( \boldsymbol{b} \) and slip plane normal \( \boldsymbol{n} \). The corresponding elastic distortion field, \(\boldsymbol{\beta}\), necessarily introduced by the dislocation was derived from analytical displacement field solutions provided by \citep{HirthLothe1992}. At the dislocation core, the  \(\boldsymbol{\beta}\) field diverges, and consequently the \(\beta_{11}\) (B) and \(\beta_{12}\) (C) fields are displayed with highly saturated color-maps making the elastic distortion variation visible far away from the dislocation core. The projected line direction \( \boldsymbol{\Bar{t}} \), Burgers vector \( \boldsymbol{\Bar{b}} \) and slip plane normal \( \boldsymbol{\Bar{n}} \) unto the \(x-y-\)plane are illustrated in (B) and (C).}
    \label{fig:sample}
\end{figure}

An aluminum crystal featuring a single straight edge dislocation was used for numerical validation. Following \cite{Borgi2024} and \cite{Poulsen2021} the Burgers vector, \( \boldsymbol{b} \), was defined along the \([1\Bar{1}0] \) direction with magnitude \(||\boldsymbol{b}||_2 = 2.86 \)\AA, the slip plane normal, \( \boldsymbol{n} \), was chosen as \( [11\Bar{1}] \), and the dislocation line direction, \( \boldsymbol{t} \), was taken along \( [112] \). The (undeformed) lattice constant was taken to be \(a=4.0493 \)\AA. The corresponding deformation gradient tensor field was derived from analytical displacement field solutions provided by \cite{HirthLothe1992} (see \ref{appendix:Deformation field surrounding an edge dislocation}). The resulting elastic distortion field is shown in Figure \ref{fig:sample}.
As shown in Figure \ref{fig:sample} the sample voxel size was taken to be 37.878 nm, with the full sample volume grid consisting of 265, 265, and 27 voxels in the \(\boldsymbol{\hat{x}}_l\), \(\boldsymbol{\hat{y}}_l\), and \(\boldsymbol{\hat{z}}_l\) dimensions, respectively. The voxel grid was aligned with the laboratory coordinate system such that the voxel face normals were parallel to the \(\boldsymbol{\hat{x}}_l\), \(\boldsymbol{\hat{y}}_l\), and \(\boldsymbol{\hat{z}}_l\)-axes, respectively. The reduced number of voxels (27 vs 265) in the z dimension was selected such that the sample covered a minimum of 4.5 standard deviations of the considered Gaussian beam profile at all (three) goniometer z-translations. Reducing the number of voxels in this way is computationally beneficial for generating synthetic DFXM images.

To facilitate the sampling outlined in section \ref{subsec:Diffraction vector measurement and acquisition geometry}, the crystal was aligned with the Cartesian lab coordinate axes such that the \([001]\) direction was along the \(\boldsymbol{\hat{z}}_l\)-axis, the \([010]\) direction along the \(\boldsymbol{\hat{y}}_l\)-axis, and the \([100]\) direction along the \(\boldsymbol{\hat{x}}_l\)-axis. This alignment ensured that the \((\Bar{1}\Bar{1}3)\), \((\Bar{1}13)\), \((113)\), and \((1\Bar{1}3)\) reflections were available for diffraction using only the \(\omega\) goniometer degree of freedom. Following section \ref{subsec:Diffraction vector measurement and acquisition geometry} the corresponding \(\omega\)-angle values for the goniometer were calculated to be \(6.43^\circ\), \(96.43^\circ\), \(186.43^\circ\), and \(276.43^\circ\). The corresponding Bragg angle for the used reflection set was \(\theta =15.416^\circ\). The azimuthal angle of diffraction was found to be \(\eta = 20.233^\circ\).

\subsection{Microscope setup \& parameters}\label{sec:Setup}
The simulated objective consisted of 69 lenses, each separated by 1600 \(\mu\)m, with a lens radius of curvature at the apex of 50 \(\mu\)m . The X-ray energy was 19.1 keV and sample to CRL distance was set to \(d_1=37.826\)cm with the corresponding image plane of the detector found at a distance \(d_2=650.899\)cm from the CRL (see Figure \ref{fig:dfxm}). This resulted in a magnification of \(\mathcal{M}\)=15.1. 

The X-ray beam profile was taken constant over \(\boldsymbol{\hat{x}}_l\) and \(\boldsymbol{\hat{y}}_l\) while following a Gaussian decay in \(\boldsymbol{\hat{z}}_l\) with a full width at half maximum (FWHM) of \(236\) nm. This is to be compared to the state-of-the-art focusing optics at the ESRF ID03 beamline which currently features a FWHM of \(\sim\)500 nm. The Gaussian resolution function described in \ref{appendix:resolution} was used and the FWHM of the CRL acceptance was taken as \(0.556\) mrad. The horizontal divergence of the beam was taken to be zero while the vertical divergence at FWHM was taken to be 0.027 mrad (similar to \cite{Poulsen2018, Poulsen2021}). The unit-less energy bandwidth (\(\Delta k/k\), see \ref{appendix:resolution}) was taken to have a standard deviation of 6 \(\times\) \(10^{-5}\). Explicitly, the covariance matrix of, \(\boldsymbol{X}\), in the stochastic model of \ref{appendix:resolution} was set to 
\begin{equation}
    \boldsymbol{\Sigma}_x = \begin{bmatrix}
        0.6469 & 0      & 0      & 0      & 0      \\
        0      & \rho      & 0      & 0      & 0      \\
        0      & 0      & 0.1315 & 0      & 0      \\
        0      & 0      & 0      & 55.7486 & 0      \\
        0      & 0      & 0      & 0      & 55.7486
        \end{bmatrix} \times 10^9,
\end{equation}
where \(\rho\) was set by the machine precision (\(< 10^{-15}\)) ensuring the existence of \(\boldsymbol{\Sigma}^{-1}_x\).

The detector was mounted with surface normal parallel to the diffracted X-ray beam and featured a pixel size of 0.75 \(\mu\)m with a total extent of 272 rows and 272 columns. The exposure time, \(t\), was calibrated separately for each of the (\(\Bar{1}\Bar{1}3\)), (\(\Bar{1}13\)), (\(113\)), and (\(1\Bar{1}3\)) reflections, ensuring that close to the full dynamic range of the camera was used (unsigned 16-bit integers, i.e the maximum value taken over all detector images was close to 65536 counts). Shot noise (Poisson) was added to each acquired detector image after which an additional thermal noise (Normal) was added. The thermal noise model was adapted to incorporate the fixed counting offset of the detector with the mean \(\mu_{thermal}=\)99.453 counts and the standard deviation \(\sigma_{thermal}=\)2.317 counts. The noise model was based on analysis of background images from a pco far-field camera (Optique Peter Twin Mic with pco.edge sCMOS camera) acquired during a recent experiment at ID03, ESRF. Prior to the addition of noise, to simulate the effect of the detector point spread function, each detector image was convoluted with a Gaussian 9 \(\times\) 9 pixel kernel featuring a standard deviation of 1 pixel. For details on state-of-the art parameters c.f \cite{Isern2024}.

 For each reflection the sample was positioned at \(z_l= \pm  \)37.878 nm and \(z_l = 0\) nm collecting data at three consecutive layers. The angular scan range was calibrated separately for each reflection resulting in the integrated angular intensity profiles found in \ref{appendix:Angular Diffraction Response}. To mimic state of the art DFXM hardware implementations the goniometer \(\phi\) and \(\chi\) motors were simulated to be positions on top of the base \(\omega\) motor. Such implementations result in non-uniform angular sampling with angular scan ranges altering in \(\phi\) and \(\chi\) between reflections. The selected scan ranges for each reflection are given in table \ref{tab:scan_ranges}.
\begin{table}[H]
\centering
\begin{tabular}{c l l l l}
\hline
Reflection (hkl) & \(\Delta\theta_{\text{range}}\) [mrad] & \(\phi_{\text{range}}\) [mrad] & \(\chi_{\text{range}}\) [mrad] & \(\omega\) [\(^\circ\)] \\
\hline
\(\Bar{1} \Bar{1} 3\) & \([-0.75, +0.75]\) & \([-0.35, +0.35]\) & \([-2.3, +2.3]\) & \(6.431585\) \\
\(\Bar{1} 1 3\) & \([-0.7, +0.7]\) & \([-2.3, +2.3]\) & \([-0.65, +0.65]\) & \(96.431585\) \\
\(1 1 3\) & \([-0.75, +0.75]\) & \([-0.35, +0.35]\) & \([-2.3, +2.3]\) & \(186.431585\) \\
\( 1 \Bar{1} 3\) & \([-0.7, +0.7]\) & \([-2.3, +2.3]\) & \([-0.65, +0.65]\) & \(276.431585\) \\
\hline
\end{tabular}
\label{tab:scan_ranges}
\caption{Angular scan ranges of the goniometer \(\phi\) and \(\chi\) motors, as well as the CRL Bragg angle increment used for each of the four simulated reflections. Note that the goniometer \(\phi\) and \(\chi\) motors are simulated to be positions on top of the base \(\omega\) motor resulting in the angular scan ranges altering in \(\phi\) and \(\chi\) between reflections.}
\end{table}
The number of angular steps in \(\phi\) and \(\chi\) was fixed to 41 for all reflections while the number CRL increments in \(\theta\) was set to 11, this resulted in a total of \(11 \times 41 \times 41\) diffraction images per z-step and reflection. Considering three sample translations in \(z\) and four reflections the simulated dataset consisted of  \(3  \times 4  \times 11 \times 41 \times 41 \approx \)200 000 2D detector pixel images.

\subsection{Deformation gradient tensor field reconstructions}\label{sec:Reconstructions}
\begin{figure}[H]
    \centering
    \includegraphics[width=0.75\linewidth]{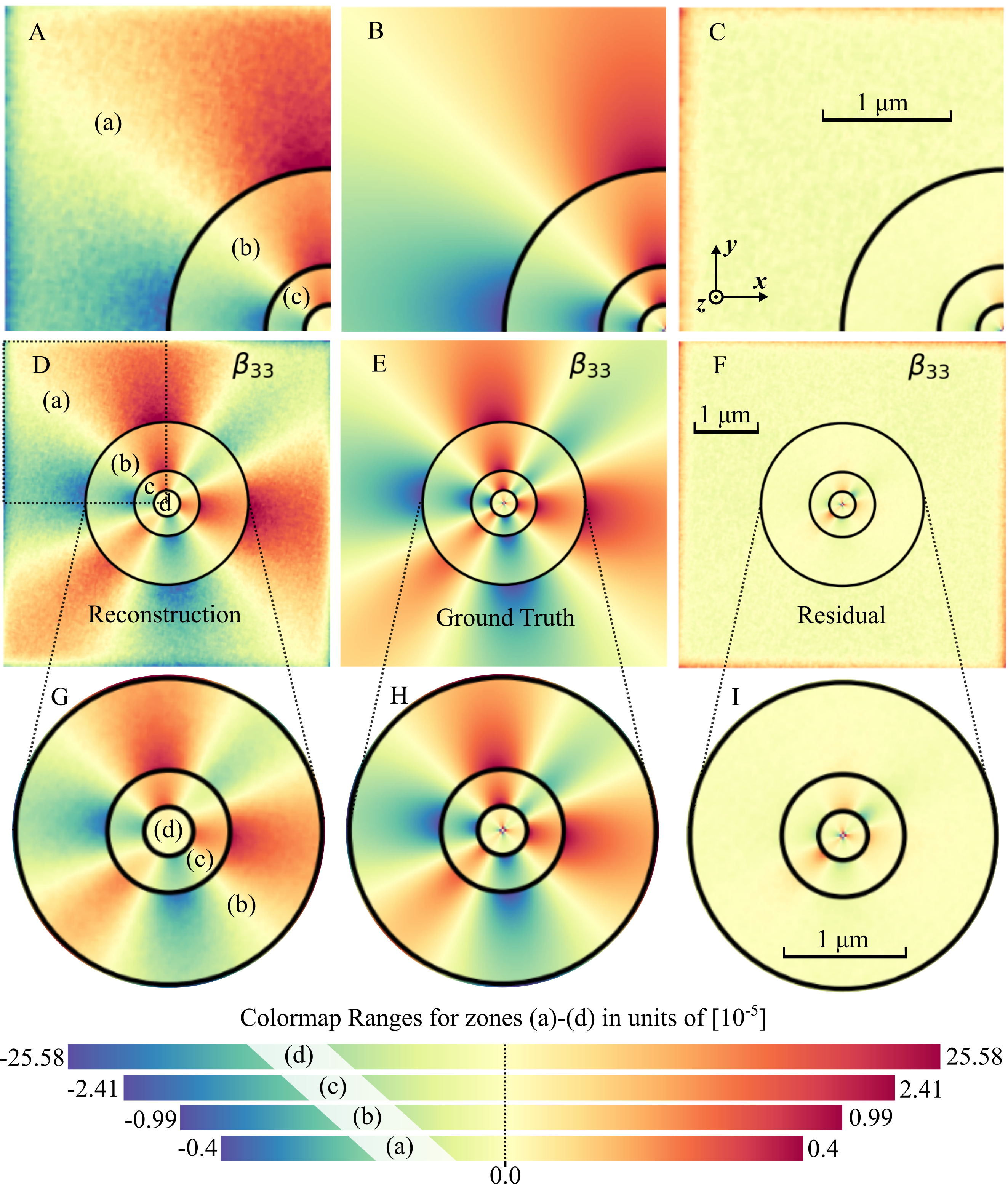}
    \caption{Reconstructed \(\beta_{33}\) component of the elastic distortion field (A,D and G) to be compared to the ground truth input elastic distortion field (B,E and H) with the difference residual shown in (C,F and I). In figures D-F an x-y slice through the center of the aluminum sample is shown. To enable simultaneous visualization of the deformation field close to and far away from the dislocation the color-map has been rescaled for each of the four zones (a)-(d) as indicated by the four color bars. Zoom in on the long ranging deformation field present in zone (a) is shown in the top row (A,B and C) while the bottom row (G,H and I) represents a zoom in on the the short ranging deformation, close to the core, (zones d,c and b). Reconstruction of the remaining 8 components of the elastic distortion tensor field can be found in \ref{appendix:(Full) Reconstructed Elastic Distortion Tensor Field} together tables showing the root mean squared error and absolute mean error for all 4 deformation zones and 9 components of the elastic distortion, \(\boldsymbol{\beta}\).}
    \label{fig:recon}
\end{figure}

\begin{figure}[H]
    \centering
    \includegraphics[width=0.95\linewidth]{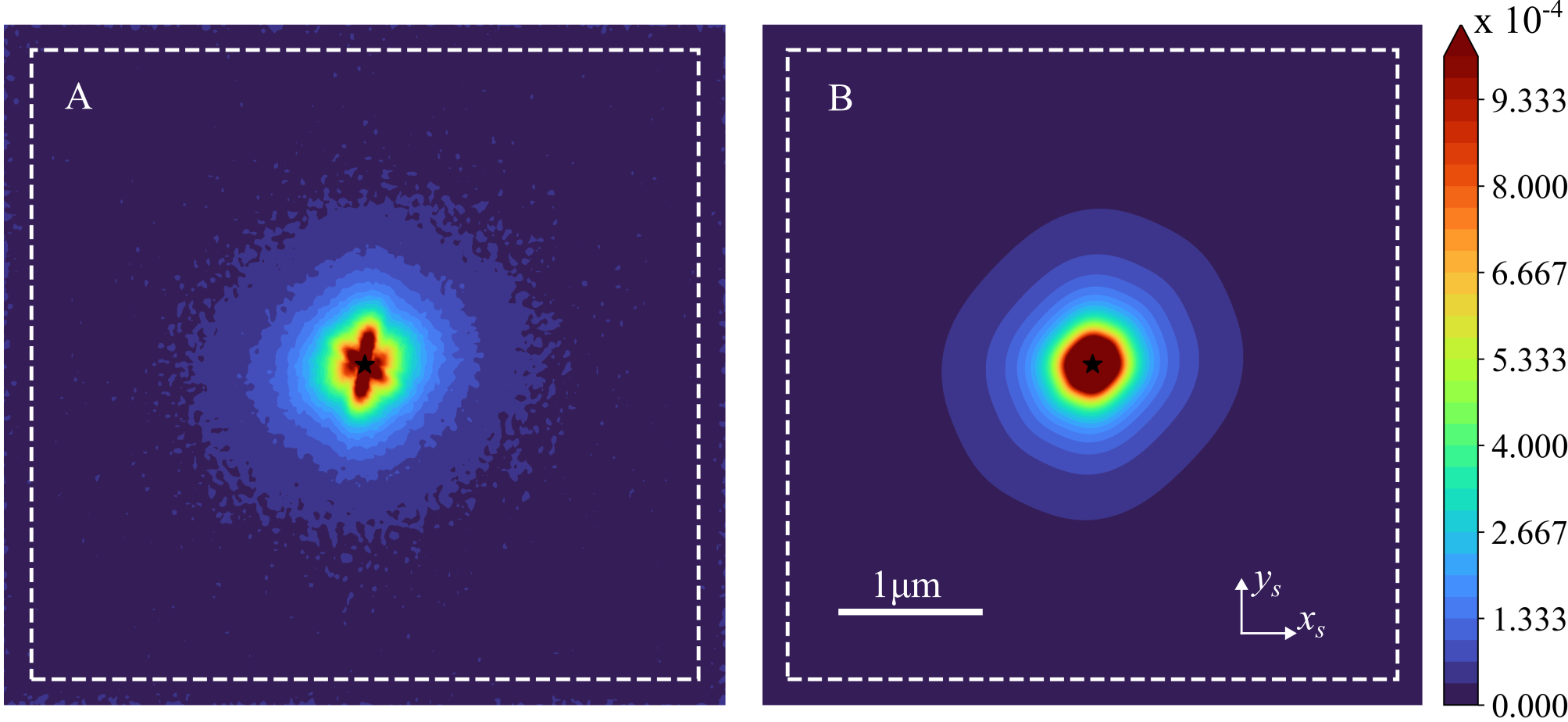}
    \caption{(A) Frobenius norm of reconstructed dislocation density tensor field. (B) Ground truth norm of dislocation density tensor field. The two black stars in (A) and (B) marks the reconstructed and true dislocation core positions respectively. The core position was reconstructed as the center of mass of \(||\boldsymbol{\alpha}||_2\) in the region enclosed by the dashed white line. The absolute error in dislocation core position was found to be 9 nm in \(x\) and 23 nm in \(y\). These values correspond to 23\% and 60\% of the voxel size  (37.88 nm) respectively. Note that the ground truth analytical \(\boldsymbol{\alpha}\) field derived through differentiation of the displacement field presented in \ref{appendix:Deformation field surrounding an edge dislocation} which sis expected to diverge in the immediate vicinity of the dislocation, this inherent artifact in the ground truth field is due to the approximations adapted by \cite{Hirth1985}.}
    \label{fig:alphanorm}
\end{figure}

\begin{figure}[H]
    \centering
    \includegraphics[width=0.7\linewidth]{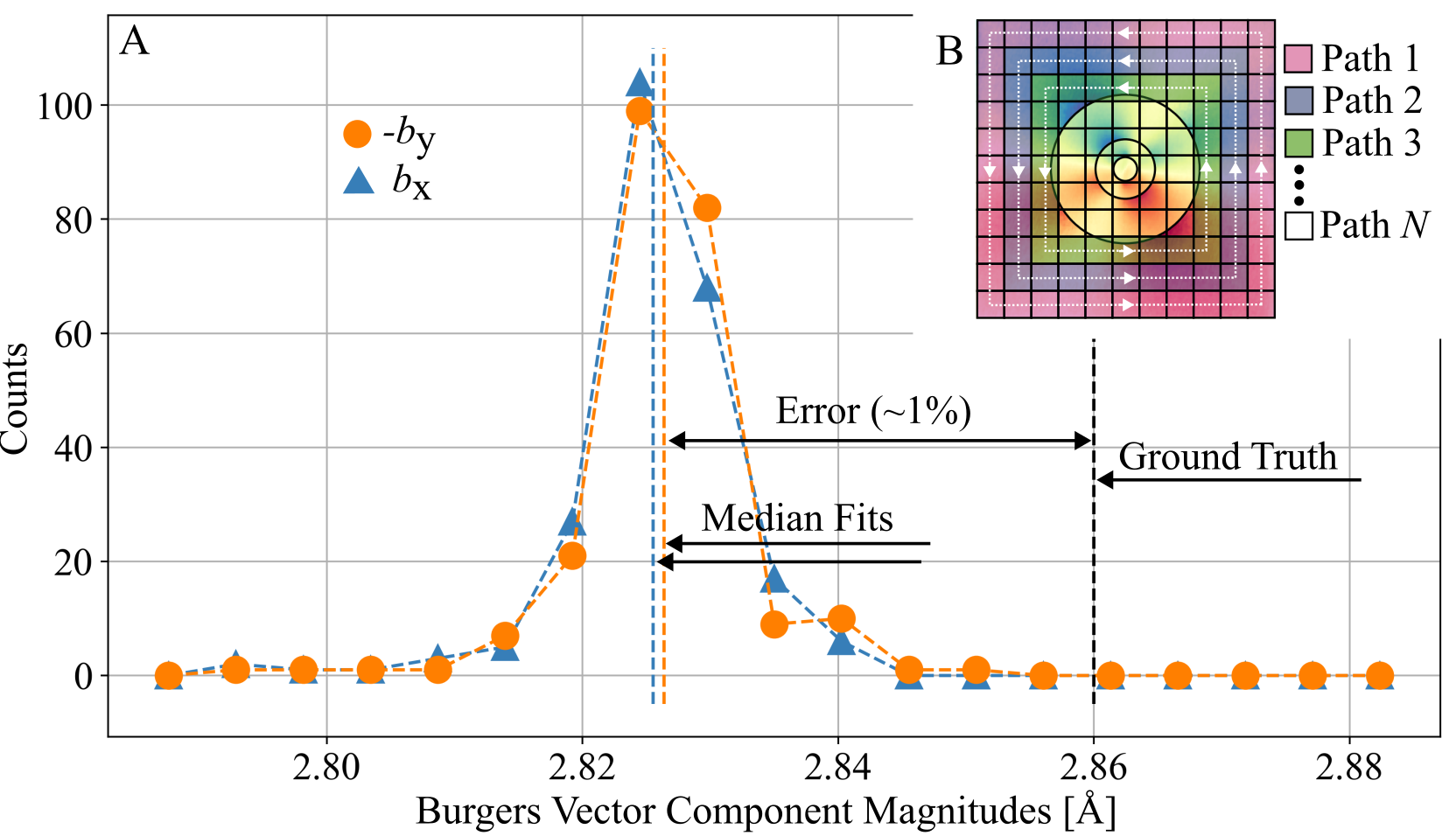}
    \caption{(A) Distribution over reconstructed non-zero Burgers vector components (\(|b_x|\), \(|b_y|\)). Each count in the histogram corresponds to a Burgers circuit integral on a loop enclosing the reconstructed dislocation core location. The total data set of the histogram corresponds to all possible square paths centered on the dislocation core and running through voxel centroids, as illustrated schematically in (B). The numerical integration was performed using the trapezoidal rule. The absolute error for the 3 Burgers vector components was \(0.0345\)\AA, \(0.0336\)\AA, and \(0.0038\)\AA for the \(x,y\) and \(z\) components respectively. Comparing to the ground truth Burgers vector, the relative error in \(b_x\) and \(b_y\) are \(1.22\%\) and \(1.19\%\) respectively.} 
    \label{fig:burger}
\end{figure}

The median detector pixel intensity value over the first five columns of the detector and all \(\phi, \chi\) and \(\Delta \theta\) positions was computed and subtracted as background. Next, for each reflection, and each \(z\)-layer, the steps outlined in Figure \ref{fig:scheme}(c-d) and detailed in section \ref{sec:CoM}-\ref{sec:Q-detector} was applied, resulting in four reconstructed \(\boldsymbol{Q}_s\) vectors per voxel in the sample. Deformation reconstruction corresponding to Figure \ref{fig:scheme}(e) and described in section \ref{sec:Local LSQ} was then applied locally for each voxel in the sample, resulting in a voxelated reconstruction of the deformation gradient tensor field \(\boldsymbol{F}_s\). The corresponding reconstructed elastic distortion field, \(\boldsymbol{\beta}_s = \boldsymbol{F}_s-\boldsymbol{I}\), is presented in Figure \ref{fig:recon} and Figure \ref{fig:fullbeta} for the central \(z\)-layer together with the ground truth elastic distortion field and the residual \(\boldsymbol{\beta}\) field, formed by subtracting the reconstructed field from the ground truth. As presented in \ref{appendix:(Full) Reconstructed Elastic Distortion Tensor Field} in tables \ref{tab:RMSE} and \ref{tab:MAE} the root mean squared error an the mean absolute error of the long-ranging residual \(\boldsymbol{\beta}\) field (zones (a) and (b) in Figure \ref{fig:recon}) was found to be in the order of \(10^{-6}\) or less for all 9 tensor components. The residual errors in zone (c), closer to the dislocation core, were found to be \(\sim\)\(10^{-5}\). The error in zone (d), in the immediate vicinity of the core, was found to be in the order of \(\sim\)\(10^{-4}\).
 From the elastic distortion presented in Figures \ref{fig:recon} and \ref{fig:fullbeta}, a reconstruction of the dislocation density tensor field was derived using equation \eqref{eq:dislocation_density} and a 2-point finite difference scheme. This transform requires spatial derivatives of \(\boldsymbol{\beta}\) to be computed along the \(x,y\) and \(z\)-directions, thus the full, 3D, reconstructed voxel volume was used (multiple \(z-\)layers). The Frobenius norm 
 \begin{equation}
    ||\boldsymbol{\alpha}||_2 = \sqrt{\sum_{ij}\alpha^2_{ij}},
\end{equation}
of the reconstructed (and true) dislocation density was then computed as presented in Figure \ref{fig:alphanorm}. From the dislocation density norm, \(||\boldsymbol{\alpha}||_2\), the core position was reconstructed with sub-pixel accuracy exploiting the fact that
\begin{equation}
    \boldsymbol{c} = \dfrac{\int_{\mathcal{S}} \boldsymbol{x} ||\boldsymbol{\alpha}(\boldsymbol{x})||_2 dx dy}{\int_{\mathcal{S}}  ||\boldsymbol{\alpha}(\boldsymbol{x})||_2 dx dy},
\end{equation}
where \(\boldsymbol{c}\) is the core position (i.e the piercing point between the dislocation line and the \(x\)-\(y\)-plane, here taken as \(\boldsymbol{0}\) in the ground truth simulation) and \(\mathcal{S}\) is any surface that symmetrically encloses the dislocation core. Here, \(\mathcal{S}\) was taken approximately by centering around the voxel with local maximum in \(||\boldsymbol{\alpha}(\boldsymbol{x})||_2\), yielding a reconstruction of the dislocation core position within a precision of 23 nm in \(x\) and \(y\).
Finally, the Burgers vector was reconstructed by integrating the elastic distortion field following \cite{ElAzab2020}. Explicitly, we have that
\begin{equation}
    \oint_{\mathcal{C}} d\boldsymbol{\ell}^T \beta(\boldsymbol{x}) = \boldsymbol{b}^{\mathcal{C}},
    \label{eq:burgers_loop}
\end{equation}
where \(\mathcal{C}\) is a loop enclosing the dislocation core, \(d\boldsymbol{l}\) is a line increment along the loop, and \(\boldsymbol{b}^{\mathcal{C}}\) is the associated Burgers vector reconstruction. We note that while it is also possible to integrate \(\boldsymbol{\alpha}\) field over the surface enclosed by the curve \(\mathcal{C}\) to recover \(\boldsymbol{b}^{\mathcal{C}}\) as described by \cite{ElAzab2020}, equation \eqref{eq:burgers_loop} is less sensitive to reconstruction artifacts existing close to the dislocation core. We used the trapezoidal integration rule and repeated the integration in \eqref{eq:burgers_loop} for every possible quadratic loop with sides parallel to the \(\boldsymbol{\hat{x}}_s\) and \(\boldsymbol{\hat{y}}_s\) axes (see Figure \ref{fig:burger}B). The resulting distribution over reconstructed Burgers vector components are illustrated in Figure \ref{fig:burger}A, the relative discrepancy between the mean reconstructed Burgers vector and the true Burgers vector was found to be less than 2\%.

\section{Discussion and Conclusions}\label{sec:Discussion}

Reconstructions of the elastic distortion field presented in Figure \ref{fig:recon} (and Figure \ref{fig:fullbeta}) and the associated Burgers vector in Figure \ref{fig:burger}, featuring a relative error of less than 2\%, validate that our method is numerically well suited to reconstruct the deformation gradient tensor field in crystalline materials.

Considering that the dislocation density tensor, \(\boldsymbol{\alpha}\), is derived from a linear combination of gradients of the elastic distortion, see equation \eqref{eq:dislocation_density}, we expect errors to multiply in regions where the \(\boldsymbol{\beta}\) field has a large residual. This is due to the fact that this equation is implemented by numerical differentiation. Evidently, this is the case in Figure \ref{fig:alphanorm}A, where, the norm of the dislocation density tensor field features strong artifacts close to the dislocation core. Indeed, in this region the input displacement field (from \cite{HirthLothe1992}) diverges, and the reconstruction presented in Figure \ref{fig:recon} fails. On the other hand, in the intermediate zones, the reconstructed beta field (\ref{fig:recon}A,D \& G) agrees well with the ground truth (\ref{fig:recon}B,E \& H) with residual errors in \(\boldsymbol{\beta}\) in the order of \(10^{-6}\) or less (see also appendix \ref{appendix:(Full) Reconstructed Elastic Distortion Tensor Field} for error quantification). Consequently, the final reconstructed core position in Figure \ref{fig:alphanorm}A was found to agree with its expected value to sub-pixel accuracy. 

The reconstruction of \(\boldsymbol{\beta}\) in zone (d) of Figure \ref{fig:recon} corresponds to the extremely weak beam conditions. The limiting factors for the quality of the reconstruction are here (I) poor signal-to-noise ratio and (II) limited angular resolution and range. Both of these challenges can, in principle, be overcome by accepting longer measurement times. Specifically, by increasing the exposure time the signal to noise ratio is improved (I) and, secondly, by expanding the scanned angular range of the goniometer and CRL while, at the same time, using a smaller angular step size, the gradient and extent of the deformation field is better sampled (II). In this way, the phantom we have selected to use in this work illustrates, for a fixed setup, the relationship between the magnitude of the gradient of deformation and the achieved accuracy of the reconstruction.

We stress that the framework derived in this paper is deformation agnostic in the sense that the underlying cause for the spatial variation in \(\boldsymbol{F}\) is arbitrary. While we have selected to present our regression scheme for a phantom that features a single dislocation line, one may equally envision the reconstruction of \(\boldsymbol{F}\) in the presence of dislocation networks, externally applied load, thermal strain, chemical strain, etc. The key limitation of our method is therefore not the origin of the spatial variation in \(\boldsymbol{F}\), but rather the relationship between the magnitude of the deformation gradient and the scanned angular ranges and step-sizes.

Building on previous published works \citep{Poulsen2017, Poulsen2021} the image formation model used in this work adapts a geometrical optics approach to DFXM. Such models entail a kinematical scattering approximation in which Bragg's law (for a perfect and infinite crystal) is assumed to hold for the infinitesimal sub-volumes of the sample. Recent work by \cite{Borgi2024} verifies that such approximations are well justified in the weak beam limit. Nevertheless, the work of \cite{Carlsen2022}, in which a dynamical diffraction model for DFXM is implemented through the use of wave propagation techniques, show much promise, especially for cases of weakly deformed crystals. The impact of a dynamical versus kinematic diffraction approximation to the final reconstructed deformation field is out of scope of the current work. For applications where an extension to dynamical diffraction is necessary, we speculate that the schemes presented in section \ref{subsec:Diffraction vector measurement and acquisition geometry} as well as the reconstruction steps in sections \ref{sec:Backpropagation} and \ref{sec:Local LSQ} will remain unchanged while the diffraction vector reconstruction step, described in section \ref{sec:Q-detector}, will likely need to be adapted.

Beyond dynamical diffraction considerations, for large sample, the effect of spatial artifacts arising from the offset of the diffraction source from the optical axis of the microscope remain to be investigated. These artifacts can manifest as systematic strain gradients in the deformation field (c.f \cite{Poulsen2017} section 4.3) and are due to the fact that the mean of the reciprocal resolution function systematically shifts within the sample volume as a function of distance to the optical axis. The incorporation of such shifts into the presented numerical framework is left for future studies. 

It is interesting to note that the framework used in this work does not feature any angular integration during strain-mosa-scanning - a procedure which is experimentally promoted in DFXM scanning. Indeed, it is typical to let one of the goniometer motors run continuously during exposure. The fact that we can recover the deformation field in Figure \ref{fig:recon} (and Figure \ref{fig:fullbeta}) despite of this lack of integration is attributed to the instrumental (reciprocal) resolution function of the instrument and the fact that the selected angular scan steps in our simulations were taken smaller than the FWHM in each of the corresponding dimensions. It is therefore likely that the results presented in section \ref{sec:Reconstructions} could be further improved by integrating continuously during exposure.

The numerical framework presented in this work was implemented in python and is planned to be packaged and open sourced released within the near future. Using a Dell Precision laptop, the total compute time to model 200 000 diffraction images, each of dimension 272 \(\times\) 272 pixels, and to reconstruct the deformation fields presented in section \ref{sec:Numerical Example} was 3 hours and 12 minutes. The vast majority of this time was consumed by the synthetic generation of DFXM images (forward modeling), which utilized two Intel(R) Core(TM) Ultra 7 155H CPUs and an NVIDIA RTX 1000 Ada Generation GPU. In contrast, the regression step, executed purely on CPUs, accounted for less than 0.1\% of the total time (i.e., under 60 seconds). The efficiency of the regression scheme arises from the data reduction step, where the mean diffraction angles in \(\phi, \chi\) and \( \Delta\theta\) are extracted from large DFXM detector image stacks, significantly reducing the data volume required for subsequent analysis. Considering that this transformation is commonly performed online during DFXM experiments, and that our regression analysis is dominated by this computation, it follows that the regression methods outlined in this paper are computationally feasible for real-time deployment in DFXM experiments. On the other hand, the reduction in dimensionality of the dataset implies that the extraction of mean diffraction angles in \(\phi, \chi\) and \( \Delta\theta\) is a type of lossy compression, in which information about the deformation gradient tensor field is lost. When computationally affordable, equation \eqref{eq:EQ_general} could be directly applied to the angular distribution of scanned intensity, maintaining, for each voxel in the sample, without loss of dimensionality, a 3D distribution over diffraction vectors, \(\boldsymbol{Q}\). Such an approach would require a treatment of the de-convolution of the microscope reciprocal resolution function with the angular distribution of scanned intensity. Overcoming this apparent challenge, however, would entail a richer inverse problem, in which a distribution over deformation gradient tensor fields is recovered. While this perspective is outside the scope of this work, it, nevertheless, merits both experimental and theoretical investigation.

In conclusion, operating under the kinematical scattering approximation, we have demonstrated, for the first time, that recovery of the full 3D deformation gradient tensor field is possible in DFXM. Our results opens the door for interfacing DFXM with CDD and DDD models and are expected to extend their applicability to larger strain ranges.

\section*{Acknowledgments}
This work was supported by the European Research Council through the Advanced Grant No. 885022 and the ESS Lighthouse on Hard Materials in Three Dimensions, SOLID.

\newpage
\appendix

\label{appendix:Searching for Oblique Reflections}
\section{Searching for Oblique Reflections}
For a given crystal symmetry, described by the lattice cell matrix \(\boldsymbol{C}_s\), and a given set of Miller planes, described by, \(\boldsymbol{Q}_{hkl}\), the task is to compute the goniometer rotation angle \(\omega\) at which the Laue condition given in equation \eqref{eq:laue} is meet. Moreover, whenever solutions exists, they are necessarily dependent on the X-ray wavelength, and, importantly, the selection of the \(\omega\)-rotation axis. In the context of crystal diffraction simulations \cite{xrdsimulator} derived general solutions, valid for arbitrary selections of \(\omega\)-rotation axis. Here, we reiterate these results for the specific rotation axis selection of \(\boldsymbol{\hat{z}}_l\) and adapt the notation to be consistent with the presented DFXM formalism. A similar treatment, specific to the DFXM setup, has also been given in \cite{Carsten2025}. 

In the following we operate in the mode where \(\phi=\chi=\mu=0\) and let the only remaining degree of freedom be \(\omega\). Once the \(\omega\)-angle of diffraction has been found, the local diffraction response can be mapped using a combination of the remaining motors \(\phi, \chi, \mu\). \cite{xrdsimulator} noted that the diffraction vector can be parameterised by a scalar value (here this scalar is directly selected as \(\omega\) while in the original publication a unit less time \(t\) was selected), adapting this approach we find that
\begin{equation}
    \boldsymbol{Q}_{l}(\omega) = 2\pi\boldsymbol{R}_{\omega}(\omega)\boldsymbol{C}_s^{-T}\boldsymbol{Q}_{hkl},
    \label{eq:omega_dependent_laue}
\end{equation}
where \(\boldsymbol{R}_{\omega}=\boldsymbol{R}_{\omega}(\omega)\) is a function of \(\omega\). The task is then to find all \(\omega\) such that equation \eqref{eq:omega_dependent_laue} is satisfied. Following \cite{xrdsimulator} we derive solutions to \eqref{eq:omega_dependent_laue} in \(\omega\) by introducing a scalar form of the Laue condition. From the elastic scattering conditions it can be shown that
\begin{equation}
\boldsymbol{k}_{l}^T\boldsymbol{Q}_{l}(\omega) + \dfrac{\boldsymbol{Q}_{l}(\omega)^T\boldsymbol{Q}_{l}(\omega)}{2} = 0.
    \label{eq:weird_bragg}
\end{equation}
Introducing \(\boldsymbol{Q}^{(0)}_{s}=2\pi\boldsymbol{C}_s^{-T}\boldsymbol{Q}_{hkl}\) and using Rodriguez rotation formula we find that
\begin{equation}
\begin{split}
     &\boldsymbol{k}_{l}^T(\boldsymbol{I}+\boldsymbol{K}^2)\boldsymbol{Q}^{(0)}_{s} + \sin( \omega)\boldsymbol{k}_{l}^T\boldsymbol{K}\boldsymbol{Q}^{(0)}_{s} - \cos( \omega)\boldsymbol{k}_{l}^T\boldsymbol{K}^2\boldsymbol{Q}^{(0)}_{s} +\dfrac{\boldsymbol{Q}^{(0)T}_{s}\boldsymbol{Q}^{(0)}_{s}}{2}=0,
     \label{eq:find_t_explicit}
\end{split}
\end{equation}
where it was used that \(\boldsymbol{Q}_{l}^T(\omega)\boldsymbol{Q}_{l}(\omega)=\boldsymbol{Q}^{(0)T}_{s}\boldsymbol{Q}^{(0)}_{s}\) since \(\boldsymbol{R}_{\omega}\) is unitary. Selecting the rotation axis to be \(\boldsymbol{\hat{z}}_l\) we find that
\begin{equation}
    \boldsymbol{K} = \begin{bmatrix}
        0   & -z_l & 0 \\
        z_l & 0    & 0 \\
        0   & 0    & 0 \\
    \end{bmatrix},
\end{equation}
however, the following derivations hold for arbitrary selections of \(\boldsymbol{K}\) (i.e an arbitrary rotation axis). To simplify, we introduce the scalar coefficients
\begin{equation}
    \begin{split}
        & \rho_0 =  -\boldsymbol{k}_{l}^T\boldsymbol{K}^2\boldsymbol{Q}^{(0)}_{s},\\
        & \rho_1 =  \boldsymbol{k}_{l}^T\boldsymbol{K}\boldsymbol{Q}^{(0)}_{s},\\
        & \rho_2 =  \boldsymbol{k}_{l}^T(\boldsymbol{I}+\boldsymbol{K}^2)\boldsymbol{Q}^{(0)}_{s} + \dfrac{\boldsymbol{Q}^{(0)T}_{s}\boldsymbol{Q}^{(0)}_{s}}{2}. \\
    \end{split}
\end{equation}
By insertion of \(\rho_0,\rho_1\) and \(\rho_2\) into equation \eqref{eq:find_t_explicit} we find that
\begin{equation}
     \rho_0\cos(\omega)+\rho_1\sin(\omega) + \rho_2 = 0.
\end{equation}
Introducing a change of variables as \(s = \tan(\omega/2)\) the double-angle formula yields
\begin{equation}
     \rho_0 \dfrac{1-s^2}{1+s^2} +\rho_1\dfrac{2s}{1+s^2} + \rho_2 = 0.
     \label{eq:quadratic_equation_find_t_explicit}
\end{equation}
Equation \eqref{eq:quadratic_equation_find_t_explicit} is a quadratic equation with one, two or zero solutions. Solving for \(s\) when \(\rho_2\neq\rho_0\) we find that
\begin{equation}
    s = \dfrac{-\rho_1}{(\rho_2 - \rho_0)}\pm\sqrt{\dfrac{c^2_1}{(\rho_2 - \rho_0)^2} - \dfrac{(\rho_0 + \rho_2)}{(\rho_2 - \rho_0)}}.
    \label{eq:tan_half_angle_solutions_in_t}
\end{equation}
The special case of \(\rho_2=\rho_0\) \eqref{eq:quadratic_equation_find_t_explicit} result in
\begin{equation}
    \rho_1 s + \rho_0 = 0, 
\end{equation}
such that only a single solution, \(s= -\rho_0/\rho_1\) can be found (given that \(\rho_1\neq 0\)). Finally, the sought rotation angle, \(\omega\), in equation \eqref{eq:omega_dependent_laue} is given by inversion of the change of variables as
\begin{equation}
    \omega = 2\arctan(s).
    \label{eq:tan_half_angle_solutions_in_s}
\end{equation}
For a given crystal cell matrix \(\boldsymbol{C}\) it is therefore possible to iterate over different \(\boldsymbol{Q}_{hkl}\) and construct a list of \(\omega\)-angles at which the various reflections are reachable in diffraction. For each such reflection, the associated \(\eta\) angles can be trivially computed from \(\boldsymbol{k'}_{l}\) which is accessible as
\begin{equation}
    \boldsymbol{k'}_{l} = \boldsymbol{Q}_{l}(\omega) + \boldsymbol{k}_{l}.
\end{equation}
Explicitly,
\begin{equation}
    \eta = \arccos\big( \boldsymbol{\hat{k}}_{yz}^T\boldsymbol{\hat{z}} \big),
\end{equation}
where
\begin{equation}
    \boldsymbol{\hat{k}}_{yz} = \dfrac{\boldsymbol{k'}_{l} - \boldsymbol{\hat{x}}^T_l\boldsymbol{k'}_{l}\boldsymbol{\hat{x}}_l}{||\boldsymbol{k'}_{l} - \boldsymbol{\hat{x}}^T_l\boldsymbol{k'}_{l}\boldsymbol{\hat{x}}_l||_2}.
\end{equation}
Importantly, due to the symmetry of the lattice, when a given \(\boldsymbol{Q}_{hkl}\) is aligned with the rotation axis, reflections typically come in groups of constant \(\eta\) angles (note that while this is a physical fact it is not a requirement for our deformation reconstruction algorithm, outlined in the paper, to work). As an example, in this paper, we considered the case of a cubic Al lattice with the 001 planes aligned with \(\boldsymbol{\hat{z}}_l\), the 100 planes aligned with \(\boldsymbol{\hat{x}}_l\) and the 010 planes aligned with \(\boldsymbol{\hat{y}}_l\) as described in section \ref{sec:Numerical Example}. Considering an energy of 19.1 keV and acceptable Bragg angles in the range \( 0 ^\circ < \theta < 16.25 ^\circ \) we solved for \(\omega\) using equation \eqref{eq:tan_half_angle_solutions_in_t} and \eqref{eq:tan_half_angle_solutions_in_s}. Sorting the resulting reflections on \(\eta\)-angles yields groups of reflections of which the first and second set are presented in tables \ref{tab:reflection_set_1} and \ref{tab:reflection_set_2}. To interpret these results, note that on a 360\(^\circ\) degree rotational interval each \(\boldsymbol{Q}_{hkl}\) becomes accessible twice, hence there is a pair of solutions \(\omega_1,\eta_1\) and \(\omega_2,\eta_2\) for each reflection. For convenience a series of solutions can be selected such that \(\eta\) and \(\theta\) remain fixed. For the cubic symmetry, the reflections are seen to be situated 90\(^\circ\) apart. For a discussion on other lattice symmetries see \cite{Carsten2025}.
\begin{table}[h]
    \centering
    \begin{tabular}{r r r r r r}
        \hline
        \(\boldsymbol{Q}_{hkl}\) & \(\omega_1\) & \(\omega_2\) & \(\eta_1\) & \(\eta_2\) & \(\theta\) \\
        \hline
        \\
        \(\Bar{1}\)\(\Bar{1}\)3 & 6.432\(^\circ\) & 263.568\(^\circ\) & 20.233\(^\circ\) & -20.233\(^\circ\) & 15.417\(^\circ\) \\
        \(\Bar{1}\)13 & 96.432\(^\circ\) & 353.568\(^\circ\) & 20.233\(^\circ\) & -20.233\(^\circ\) & 15.417\(^\circ\) \\
        1\(\Bar{1}\)3 & 173.568\(^\circ\) & 276.432\(^\circ\) & -20.233\(^\circ\) & 20.233\(^\circ\) & 15.417\(^\circ\) \\
        113 & 83.568\(^\circ\) & 186.432\(^\circ\) & -20.233\(^\circ\) & 20.233\(^\circ\) & 15.417\(^\circ\) \\
    \end{tabular}
    \caption{Primary accessible reflection set (lowest absolute \(\eta\)-angle) for an X-ray energy of 19.1 keV. The four reflections can be reached at the same fixed \(\theta\) and \(\eta\) angle when \(\phi=\chi=\mu=0\) by varying only the \(\omega\)-rotation.}
    \label{tab:reflection_set_1}
\end{table}

\begin{table}[h]
    \centering
    \begin{tabular}{r r r r r r}
        \hline
        \(\boldsymbol{Q}_{hkl}\) & \(\omega_1\) & \(\omega_2\) & \(\eta_1\) & \(\eta_2\) & \(\theta\) \\
        \hline
        \\
        \(\Bar{2}\)02 & 71.300\(^\circ\) & 288.700\(^\circ\) & 43.447\(^\circ\) & -43.447\(^\circ\) & 13.103\(^\circ\) \\
        0\(\Bar{2}\)2 & 341.300\(^\circ\) & 198.700\(^\circ\) & 43.447\(^\circ\) & -43.447\(^\circ\) & 13.103\(^\circ\)\\
        022 & 161.300\(^\circ\) & 18.700\(^\circ\) & 43.447\(^\circ\) & -43.447\(^\circ\) & 13.103\(^\circ\) \\
        202 & 108.700\(^\circ\) & 251.300\(^\circ\) & -43.447\(^\circ\) & 43.447\(^\circ\)& 13.103\(^\circ\) \\

    \end{tabular}
    \caption{Secondary accessible reflection set (second lowest absolute \(\eta\)-angle) for an X-ray energy of 19.1 keV. The four reflections can be reached at the same fixed \(\theta\) and \(\eta\) angle when \(\phi=\chi=\mu=0\) by varying only the \(\omega\)-rotation.}
    \label{tab:reflection_set_2}
\end{table}

\newpage
\section{Deformation field surrounding an edge dislocation}\label{appendix:Deformation field surrounding an edge dislocation}
Using a local coordinate system \cite{HirthLothe1992} described the displacement field, \( \boldsymbol{u}_{d}(\boldsymbol{x}_d) : \mathbb{R}^{3} \to \mathbb{R}^{3} \), resulting from a single straight edge dislocation as
\begin{equation}
\begin{split}
u_{d,x} &= \frac{b}{2\pi} \left[ \arctan \left( \frac{y_d}{x_d} \right) + \frac{x_d y_d}{2(1 - \nu)(x_d^2 + y_d^2)} \right], \\
u_{d,y} &= -\frac{b}{2\pi} \left[ \frac{1 - 2\nu}{4(1 - \nu)} \ln(x_d^2 + y_d^2) + \frac{x_d^2 - y_d^2}{4(1 - \nu)(x_d^2 + y_d^2)} \right], \\
u_{d,z} &= 0,
\end{split}
\label{eq:ud}
\end{equation}
where the dislocation frame of reference is subscripted \(d\) and defined with \(\boldsymbol{\hat{z}}_d\) along the line direction, \(\boldsymbol{t}\), \(\boldsymbol{\hat{x}}_d\) along the Burgers vector, \(\boldsymbol{b}\), and \(\boldsymbol{\hat{y}}_d\) parallel to the normal of the dislocation glide planes, \(\boldsymbol{n}\). It follows that the origin of this local reference frame is situated at the dislocation core. The deformation gradient tensor become in this frame of reference as
\begin{equation}
\boldsymbol{F}_d = \boldsymbol{I} + 
\begin{bmatrix}
\frac{\partial u_{d,x}}{\partial x} & \frac{\partial u_{d,x}}{\partial y} & 0 \\
\frac{\partial u_{d,y}}{\partial x} & \frac{\partial u_{d,y}}{\partial y} & 0 \\
0 & 0 & 0
\end{bmatrix}.
\label{eq:F_d_def}
\end{equation}
Applying the partial derivatives in \eqref{eq:F_d_def} with respect to the analytical expressions in \eqref{eq:ud} we find that
\begin{equation}
\begin{split}
F^d_{11} &= 1 - c_dy_d( 3x_d^2 + y_d^2 - l_d ), \\
F^d_{12} &= c_dx_d( 3x_d^2 + y_d^2 - l_d ), \\
F^d_{21} &= -c_dx_d( x_d^2 + 3y_d^2 - l_d ), \\
F^d_{22} &= 1 + c_dy_d(x_d^2 - y_d^2 + l_d) , \\
F^d_{23} &= F^d_{32} = F^d_{31} = F^d_{13} = 0, \\
F^d_{33} &= 1,
\end{split}
\end{equation}
where
\begin{equation}
c_d = \frac{b}{ 4\pi(1 - \nu)r_d^2 }, \quad
l_d = 2\nu r_d, \quad \text{and} \quad 
r_d = x_d^2 + y_d^2.
\end{equation}
The mapping from dislocation system to crystal coordinates is described as
\begin{equation}
    \boldsymbol{x}_c = \boldsymbol{U}_d \boldsymbol{x}_d + \Delta \boldsymbol{x}_c,
\end{equation}
where \(\Delta \boldsymbol{x}_d\) is a vector from the dislocation system origin to the crystal coordinate origin (described in crystal coordinates) and
\begin{equation}
    \boldsymbol{U}_d = \begin{bmatrix}
        \boldsymbol{\hat{x}}_d & \boldsymbol{\hat{y}}_d & \boldsymbol{\hat{z}}_d
    \end{bmatrix}.
\end{equation}
The sample space deformation gradient tensor therefore become
\begin{equation}
    \boldsymbol{F}_s =  \boldsymbol{U} \boldsymbol{U}_d \boldsymbol{F}_d \boldsymbol{U}^T_d \boldsymbol{U}^T.
\end{equation}
See also \cite{Poulsen2021} section 8.1 for a comparison. 

\newpage
\section{Scan patterns for the \(\Bar{1}\Bar{1}3\), \(\Bar{1}13\), \(113\), \(1\Bar{1}3\) reflections}\label{appendix:Angular Diffraction Response}
\begin{figure}[H]
    \centering
    \includegraphics[scale=0.8]{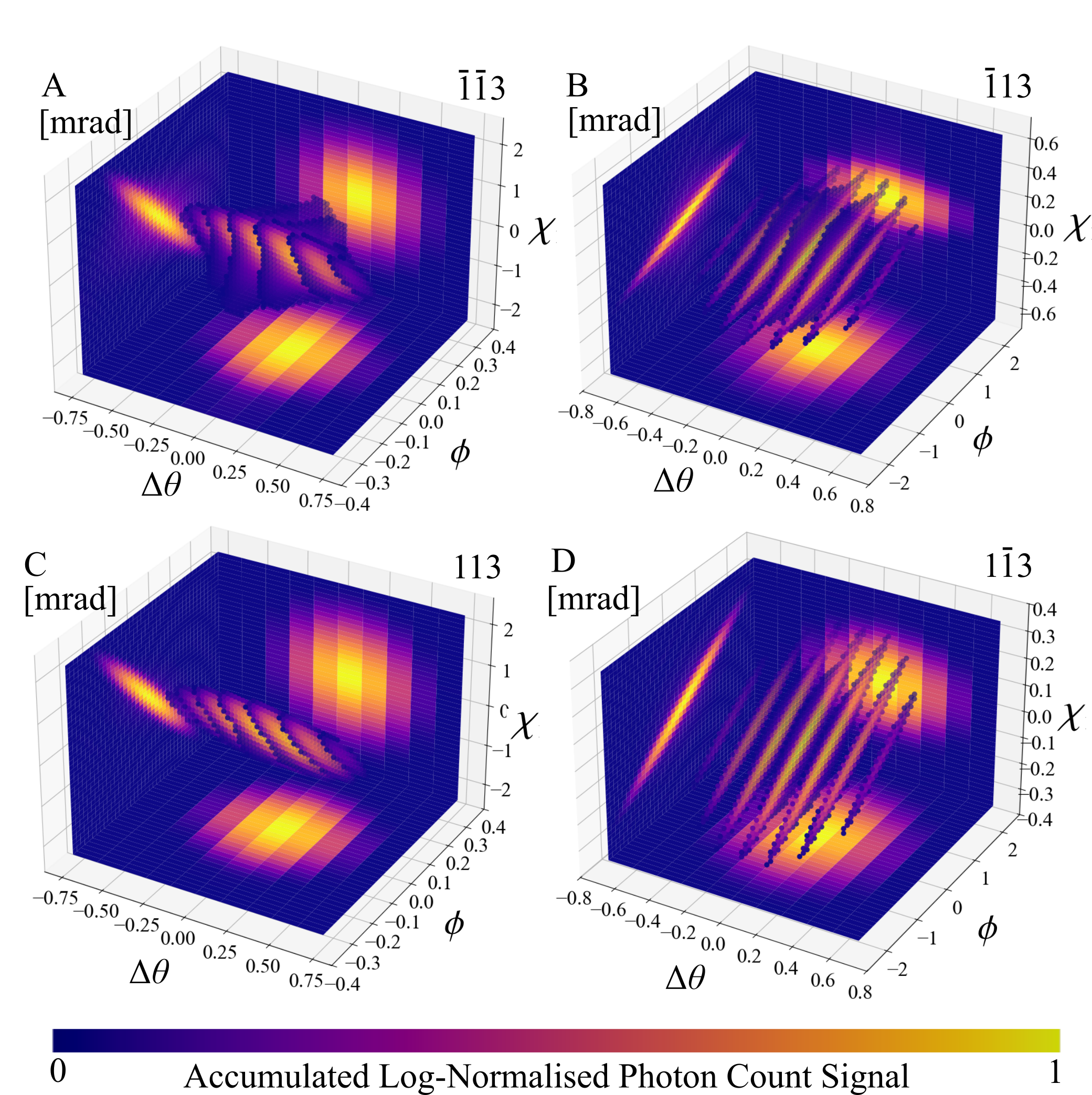}
    \caption{Accumulated log-normalized diffraction response for the four reflections used in the simulation. The 3D scatter plots in A-D corresponds to the photon count values summed over the two detector dimensions \(u\) and \(v\). The resulting intensity distributions are referred to as a "scan patterns". For better visualization the scan patterns have been summed along the Cartesian axes, re-normalized, and plotted as 2D intensity profile projections. The mesh of the Cartesian projections highlight the nonuniform angular step size which features the largest angular step in the \(\Delta \theta\) direction.}
    \label{fig:scanpatterns}
\end{figure}

For any angular setting of  \(\Delta \theta\), \(\phi\) and \(\chi\) a DFXM image is collected by the detector in our simulation. This 2D image of diffraction naturally varies in intensity from pixel to pixel due to the sample deformation and due to noise. Additionally there is a diffraction response variation between frames which is dominated by the sample deformation. Therefore, to visualize the angular support of the sample, we introduce the so called "scan pattern function", \(f(\Delta \theta, \phi, \chi)\), as the continuous function over angular space \((\Delta \theta, \phi, \chi)\) with scalar value equal to the integrated intensity over the detector area. The shape of the scan patter function, \(f\), will depend jointly on the microscope resolution function and the sample deformation field. The support of \(f\) will, in general, determine the magnitude of the angular ranges that need be scanned to achieve a complete sampling of the deformation field. To achieve such angular coverage, we tuned the angular scan ranges for the four used reflections (\(\Bar{1}\Bar{1}3\)), (\(\Bar{1}13\)), (\(113\)) and (\(1\Bar{1}3\)) which resulted in the scan patterns illustrated in Figure \ref{fig:scanpatterns}. We note that in the search for the support of \(f\) a coarser angular grid can be beneficial. We also note that a similar angular sweeping procedure as described here is typically deployed at ID03 ESRF to locate the sample in angular space.

\newpage
\section{(Full) Reconstructed Elastic Distortion Tensor Field}
\label{appendix:(Full) Reconstructed Elastic Distortion Tensor Field}
\begin{figure}[H]
    \centering
    \includegraphics[scale=0.8]{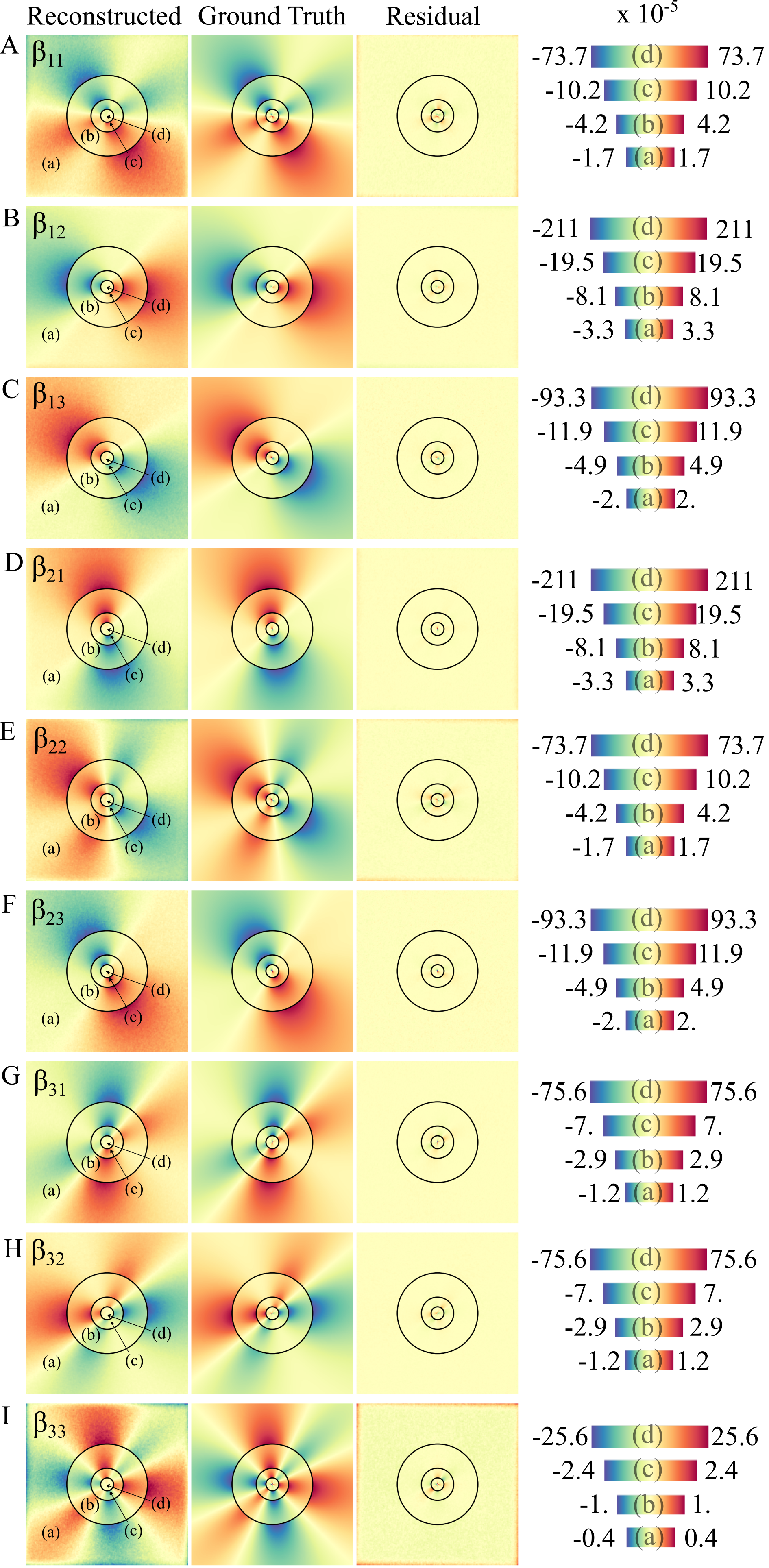}
    \caption{Reconstructed (left column) True (middle column) and residual (right column) elastic distortion tensor field for the middle \(z\)-layer in the aluminum sample described in section \ref{sec:Numerical Example}. For visual purposes the color bar range has been rescaled in the four regions a-d. The \(\boldsymbol{\beta}_{ij}\) components are displayed as: A-\(\beta_{11}\), B-\(\beta_{12}\), C-\(\beta_{13}\),  D-\(\beta_{21}\), E-\(\beta_{22}\), F-\(\beta_{23}\),  G-\(\beta_{31}\), H-\(\beta_{32}\), I-\(\beta_{33}\), in units of \(10^{-5}\)}
    \label{fig:fullbeta}
\end{figure}
To quantify the error in the reconstructions given in Figure \ref{fig:fullbeta} we present, in tables \ref{tab:MAE} and \ref{tab:RMSE}, the root mean squared error and the mean absolute error of the residual \(\boldsymbol{\beta}\) field corresponding to the rightmost column of Figure \ref{fig:fullbeta}. The definition of the root mean squared error, \(RMSE\), of the component \(\beta_{ij}\) is defined as
\begin{equation}
    RMSE = \sqrt{\dfrac{\sum^{k=n}_{k=1} (\beta^r_{k,ij} - \beta^t_{k,ij})^2}{n}}
    \label{eq:RMSE}
\end{equation}
where \(n\) is the number of considered voxels, \(\beta^t_{k,ij}\) is the ground truth elastic distortion at voxel number \(k\) and \(\beta^r_{k, ij}\) is the reconstructed elastic distortion at voxel number \(k\). Similarly, the mean absolute error, \(MAE\), of the component \(\beta_{ij}\) is defined as
\begin{equation}
    MAE =  \dfrac{ \sum^{k=n}_{k=1} |\beta^r_{k,ij} - \beta^t_{k,ij}| }{n}
    \label{eq:MAE}
\end{equation}

\begin{table}[h]
\centering
\caption{Root mean squared error of the residual \(\boldsymbol{\beta}\) field corresponding to the rightmost column of Figure \ref{fig:fullbeta}. The error is given per deformation zones (a-d) in the four columns and seen to be at the largest close to the dislocation core (zone d). The values are presented in units of \(\times\) \(10^{-5}\).}
\begin{tabular}{c ccc c}
\hline
           & zone (a) & zone (b) & zone (c) & zone (d) \\
\hline
$\beta_{11}$ & 0.067 & 0.122 & 1.289 & 11.814 \\
$\beta_{12}$ & 0.058 & 0.124 & 1.365 & 22.049 \\
$\beta_{13}$ & 0.037 & 0.085 & 0.928 & 12.597 \\
$\beta_{21}$ & 0.060 & 0.102 & 0.666 & 22.043 \\
$\beta_{22}$ & 0.065 & 0.142 & 1.238 & 10.837 \\
$\beta_{23}$ & 0.037 & 0.079 & 0.641 & 12.231 \\
$\beta_{31}$ & 0.012 & 0.049 & 0.537 & 8.108 \\
$\beta_{32}$ & 0.012 & 0.049 & 0.570 & 8.575 \\
$\beta_{33}$ & 0.040 & 0.042 & 0.333 & 3.892 \\
\hline
\end{tabular}
\label{tab:RMSE}
\end{table}

\begin{table}[h]
\centering
\caption{Mean absolute error the residual \(\boldsymbol{\beta}\) field corresponding to the rightmost column of Figure \ref{fig:fullbeta}. The error is given per deformation zones (a-d) in the four columns and seen to be at the largest close to the dislocation core (zone d). The values are presented in units of \(\times\) \(10^{-5}\). }
\begin{tabular}{c ccc c}
\hline
            & zone (a) & zone (b) & zone (c) & zone (d) \\
\hline
$\beta_{11}$ & 0.051 & 0.089 & 0.933 & 7.460 \\
$\beta_{12}$ & 0.038 & 0.087 & 0.983 & 10.163 \\
$\beta_{13}$ & 0.029 & 0.062 & 0.679 & 6.576 \\
$\beta_{21}$ & 0.039 & 0.074 & 0.518 & 8.672 \\
$\beta_{22}$ & 0.047 & 0.097 & 0.915 & 6.738 \\
$\beta_{23}$ & 0.029 & 0.056 & 0.455 & 5.152 \\
$\beta_{31}$ & 0.009 & 0.033 & 0.378 & 3.997 \\
$\beta_{32}$ & 0.009 & 0.034 & 0.381 & 4.297 \\
$\beta_{33}$ & 0.031 & 0.034 & 0.236 & 2.136 \\
\hline
\end{tabular}
\label{tab:MAE}
\end{table}

\newpage
\section{Errors due to Taylor expansion of \(\boldsymbol{f}\)}\label{appendix:Taylor expansion of f}
\begin{figure}[H]
    \centering
    \includegraphics[width=0.8\linewidth]{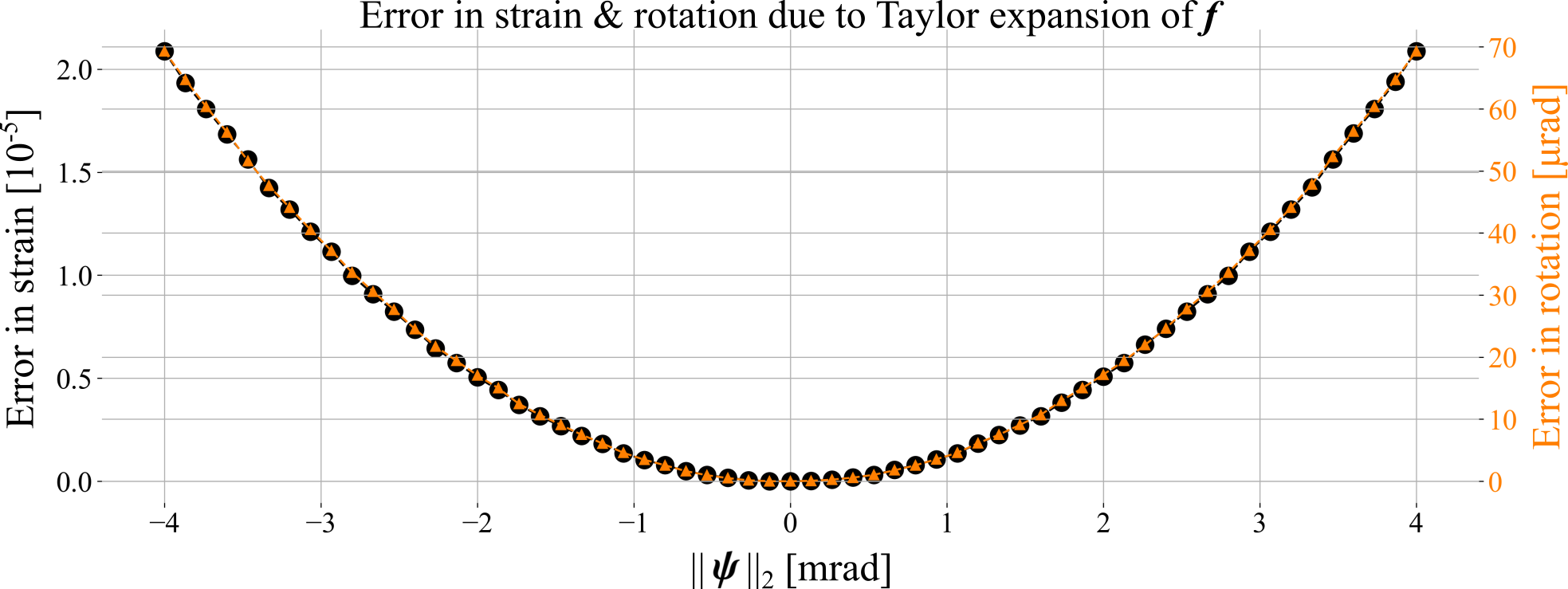}
    \caption{Error in strain and rotation as a result of Taylor expanding the map \(\boldsymbol{Q}_s = \boldsymbol{f}(\boldsymbol{\psi})\) to first order. The error is seen to increase with \(||\boldsymbol{\psi}||_2\). For DFXM this implies that the error increases with the sample magnitude of deformation. This follows from the fact that larger deformation requires larger scan ranges in \(\boldsymbol{\psi}\). The analysis corresponds to the \(\Bar{1}\Bar{1}\Bar{3}\) reflection from the Al sample used in section \ref{sec:Numerical Example} (\(\eta=20.233^\circ\), \(\theta=15.417^\circ\)). }
    \label{fig:errorraylor}
\end{figure}

The Taylor series expansion of \(\boldsymbol{f}(\boldsymbol{\psi})\) in equation \eqref{eq:taylor_f} implies an error, \(\Delta \boldsymbol{Q}\), in the right hand side integral of \eqref{eq:EQ_general}. Considering that \(p_{\Psi}\) is a probability density (integrating to unity) an upper bound to this error can be given by a
\begin{equation}
\begin{split}
    & |\Delta \boldsymbol{Q}| < \text{max}\bigg(\big|\boldsymbol{g}(\boldsymbol{\psi}) - \boldsymbol{f}(\boldsymbol{\psi})\big|\bigg) \quad \forall \quad \boldsymbol{\psi}, \\
    & \boldsymbol{g}(\boldsymbol{\psi}) = \boldsymbol{f}(\boldsymbol{0}) + \boldsymbol{\psi}^T\dfrac{\partial \boldsymbol{f}}{\partial \boldsymbol{\psi}}\bigg\rvert_{\boldsymbol{\psi}=\boldsymbol{0}}
    \label{eq:appendix:error_f}.
\end{split}
\end{equation}
In Figure \ref{fig:errorraylor} we give this bound in units of strain and misorientation respectively. The results where achieved by first evaluating \(\boldsymbol{f}\) and \(\boldsymbol{g}\) on an equidistant grid over \(\Delta\theta, \phi\) and \(\chi\) using a grid spacing of \(2.327\mu\)rad. The misorientation was defined as
\begin{equation}
    \Delta m = \arccos\bigg( \dfrac{\boldsymbol{g}^T\boldsymbol{f}}{||\boldsymbol{g}||_2 ||\boldsymbol{f}||_2} \bigg)
\end{equation}
and the error in strain was defined as
\begin{equation}
\begin{split}
    & \Delta s = \dfrac{||\boldsymbol{f}(\boldsymbol{0})||_2}{||\boldsymbol{f}||_2 } - \dfrac{||\boldsymbol{f}(\boldsymbol{0})||_2}{||\boldsymbol{g}||_2 } \\
\end{split}
\end{equation}
By converting the computed values of \(\boldsymbol{f}\) and \(\boldsymbol{g}\) values into strain and misorientation and searching over this space in accordance with \eqref{eq:appendix:error_f} on increasing spherical subdomains (increasing \(||\boldsymbol{\psi}||_2\)) the upper error bound on \(\mathbb{E}[\boldsymbol{Q}_s| \boldsymbol{u}]\), shown in Figure \ref{fig:errorraylor}, was found. The analysis corresponds to the \(\Bar{1}\Bar{1}\Bar{3}\) reflection from the Al sample used in section \ref{sec:Numerical Example} (\(\eta=20.233^\circ\), \(\theta=15.417^\circ\)).

\newpage
\section{Analytical Resolution Function}\label{appendix:resolution}
To model image formation in DFXM, the reciprocal resolution function, \(r(\boldsymbol{Q}_l)\), plays a critical role. It determines the intensity deposited at the detector for each simulated voxel in the sample, making efficient evaluation of the reciprocal space resolution function crucial for performance. In the following we give account of a new closed form multivariate Gaussian expression for the reciprocal resolution function in DFXM. The derived expressions corresponds to a previously published stochastic model that is due to \cite{Poulsen2017}. Our analytical contribution allow for the resolution function to be probed without the need for Monte Carlo sampling. These results are valid for any resolution model in which the ingoing stochastic variables are multivariate Gaussian. For non-Gaussian resolution functions c.f \cite{Poulsen2021}.

\subsection{Recapitulation of the Stochastic Model of \cite{Poulsen2017}}
In \cite{Poulsen2017}, a Gaussian model for the reciprocal space resolution function in DFXM was proposed. This model employs a small-angle approximation to linearize the perturbations in the incident and diffracted wavevectors with respect to the driving stochastic variables. Specifically, the model describes the perturbation of the incident wavevector as:
\begin{equation}
    \Delta \boldsymbol{k}_l = k
\begin{bmatrix}
    \varepsilon \\
     \varsigma_h \\
     \varsigma_v
\end{bmatrix}_{l},
\label{eq:appendix:kin}
\end{equation}
and the outgoing wavevector perturbation as:
\begin{equation}
    \Delta \boldsymbol{k'}_l = k \boldsymbol{R}_{\eta} \boldsymbol{R}_{2\theta}^T
\begin{bmatrix}
    \varepsilon \\
     \xi_h \\
     \xi_v
\end{bmatrix}_{l},
\label{eq:appendix:kout}
\end{equation}
where the driving variables are \(\varsigma_h\), \( \varsigma_v\), \( \xi_h\), and \( \xi_v\), which are all Gaussian-distributed random variables representing small perturbations in the horizontal and vertical components of the wavevectors. Likewise \(\varepsilon\) is a random Gaussian perturbation in the energy of the ray such that \(\Delta k / k = \varepsilon\). The above model leads to a Gaussian perturbation in the diffraction vector as
\begin{equation}
    \Delta \boldsymbol{Q}_l = \Delta \boldsymbol{k'}_l - \Delta \boldsymbol{k}_l,
    \label{eq:appendix:dQ}
\end{equation}
where \(\Delta\boldsymbol{Q}_l=\boldsymbol{Q}_l - \boldsymbol{Q}^{(0)}\) is a perturbation around the nominal diffraction vector \(\boldsymbol{Q}_l^{(0)}\).

\subsection{Derivation of Analytical Solutions}
Given that \(\Delta \boldsymbol{Q}_l\) is linear in the driving variables we seek an expression of the form
\begin{equation}
   \Delta \boldsymbol{Q}_l =  \boldsymbol{M}\boldsymbol{X},
\end{equation}
where
\begin{equation}
    \boldsymbol{X} = \begin{bmatrix}
        \varepsilon &
        \varsigma_h &
        \varsigma_v &
        \xi_h &
        \xi_v &
    \end{bmatrix}^T,
\end{equation}
and \(\boldsymbol{M}\in\mathbb{R}^{3 \times 5}\) is a  constant matrix. By multiplying out the expressions in \eqref{eq:appendix:kin} and \eqref{eq:appendix:kout} and inserting into equation \eqref{eq:appendix:dQ} the matrix \(\boldsymbol{M}\) can be obtained by factoring out a term \(\boldsymbol{X}\). Denoting \(c_{\eta} = \cos(\eta)\), \(s_{\eta} = \sin(\eta)\) and \(c_{\theta} = \cos(2 \theta)\), \(s_{\theta} = \sin(2 \theta)\) we find that
\begin{equation}
\boldsymbol{M}^T = k
\begin{bmatrix} c_\theta - 1 & -  s_\eta s_\theta &  s_\theta c_\eta \\0 & -1  & 0\\0 & 0 & -1 \\0 &   c_\eta&   s_\eta\\-  s_\theta & -  s_\eta c_\theta &  c_\eta c_\theta \end{bmatrix}.
\end{equation}

\begin{figure}[H]
    \centering
    \includegraphics[width=0.7\linewidth]{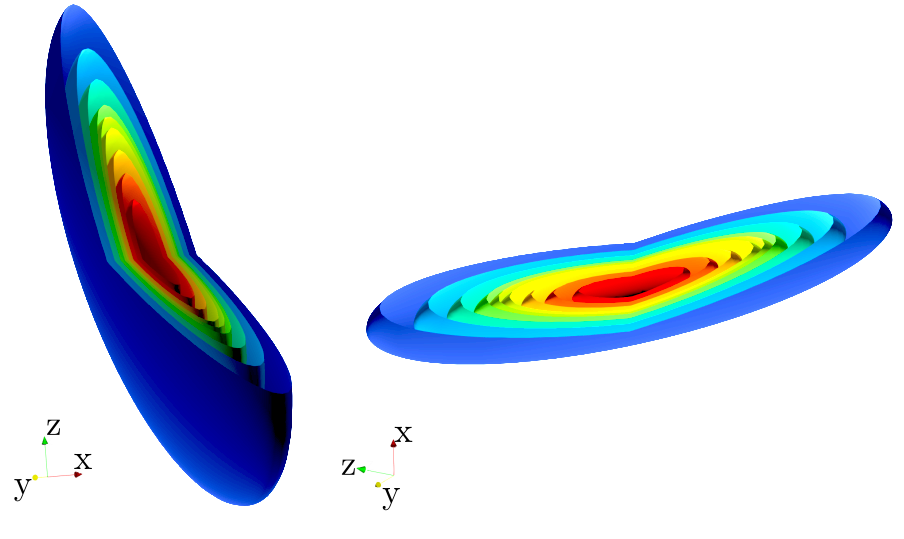}
    \caption{Schematic rendering of level sets of the resolution function derived in equation \eqref{eq:appendix:r_Ql}. The Gaussian is a thin disc corresponding to the fact that the Debye-Scherrer cone be more rapidly escaped by rotating the crystalline element in \(\phi\) as opposed to \(\chi\) (given that \(\omega=0\)).}
    \label{fig:res}
\end{figure}

Considering that \(\Delta\boldsymbol{Q}_l\) is a linear transformation of the multivariate Gaussian variable \(\boldsymbol{X}\) we find the covariance of \(\Delta\boldsymbol{Q}_l\) as
\begin{equation}
    \boldsymbol{\Sigma}_Q = \boldsymbol{M} \boldsymbol{\Sigma}_X \boldsymbol{M}^T,
\end{equation}
where the covariance of \(\boldsymbol{X}\) is denoted \(\boldsymbol{\Sigma}_X\). Similarly, letting the mean of \(\boldsymbol{X}\) be \(\boldsymbol{\mu}_X = \boldsymbol{0}\), we find that
\begin{equation}
    \boldsymbol{\mu}_Q = \boldsymbol{M} \boldsymbol{\mu}_X = \boldsymbol{0},
\end{equation}
where \(\boldsymbol{\mu}_Q =\boldsymbol{0}\) is the mean of \(\Delta\boldsymbol{Q}_l\). Thus, we may write the reciprocal resolution function explicitly as the multivariate Gaussian
\begin{equation}
    r(\Delta\boldsymbol{Q}_l) = c_Q \exp \left( - \frac{1}{2} \Delta\boldsymbol{Q}_l^T \boldsymbol{\Sigma}_Q^{-1} \Delta\boldsymbol{Q}_l \right),
    \label{eq:appendix:r}
\end{equation}
where the normalizing constant is
\begin{equation}
    c_Q = \frac{1}{\sqrt{(2\pi)^3 \det(\boldsymbol{\Sigma}_Q)}}.
\end{equation}
Alternatively, we may use the fact that \(\Delta\boldsymbol{Q}_l=\boldsymbol{Q}_l - \boldsymbol{Q}^{(0)}\) to write
\begin{equation}
    r(\Delta\boldsymbol{Q}_l) = c_Q \exp \left( - \frac{1}{2} (\boldsymbol{Q}_l - \boldsymbol{Q}^{(0)})^T \boldsymbol{\Sigma}_Q^{-1} (\boldsymbol{Q}_l - \boldsymbol{Q}^{(0)}) \right).
    \label{eq:appendix:r_Ql}
\end{equation}
The analytical expression in equation \eqref{eq:appendix:r_Ql} can be used to render the resolution function in lab coordinates, as seen in Figure \ref{fig:res}, the resulting multivariate Gaussian is typically characterized as a thin disc.

\bibliographystyle{elsarticle-harv} 
\bibliography{bib.bib}

\end{document}